\documentclass[fleqn,usenatbib]{mnras}
\usepackage{xeCJK}

\usepackage{newtxtext}
\usepackage[T1]{fontenc}
\usepackage{ae,aecompl}

\usepackage{graphicx}	% Including figure files
\usepackage{amsmath}	% Advanced maths commands
\usepackage{amssymb}	% Extra maths symbols
\usepackage[british]{babel}             % British English hyphenation
\usepackage{bm}
\usepackage{physics}
\usepackage{academicons}
\usepackage[dvipsnames]{xcolor}
\definecolor{orcidlogocol}{HTML}{A6CE39}
\newcommand{\orcid}[1]{\href{https://orcid.org/#1}{\textcolor[HTML]{A6CE39}{\aiOrcid}}}
\usepackage{booktabs}
\usepackage{makecell}
\usepackage{ulem}

\usepackage[nolist]{acronym}
\newacro{cpd}[CPD]{circumplanetary disk}
\newacro{RSDs}{redshift space distortions}
\newacro{RSD}{redshift space distortions}
\newacro{DGP}{Dvali-Gabadadze-Porrati}
\newacro{CMB}{cosmic microwave background}
\newacro{SM}{Streaming Model}
\newacro{GSM}{Gaussian Streaming Model}
\newacro{STSM}{Skew-T Streaming Model}

\newacro{LPT}{Lagrangian Perturbation Theory}
\newacro{CLPT}{Convolutional Lagrangian Perturbation Theory}
\newacro{GR}{General Relativity}
\newacro{MG}{Modified Gravity}
\newacro{ST}{skewed student-t}
\newacro{PDF}{Probability Distribution Function}
\newacro{FFT}{Fast Fourier Transform}
\newacro{HOD}{Halo Occupation Distribution}
\newacro{SHAM}{subhalo abundance matching}
\newacro{HMF}{halo mass function}
\newacro{LSS}{large-scale structure}
\newacro{TPCF}{two-point correlation function}
\newacro{EFT}{effective field theory}

\title[Halo Occupations \& Assembly Bias in MG]{Galaxy formation in modified gravity -- II. galaxy halo connection and assembly bias}

\author[M.~Collier et al.]{Michael Collier\orcid{0000-0001-5628-3837}$^{1}$\thanks{E-mail: michael.collier@durham.ac.uk}, 
Sownak Bose\orcid{0000-0002-0974-5266}$^1$\thanks{E-mail: sownak.bose@durham.ac.uk},
Baojiu Li\orcid{0000-0002-1098-9188}$^1$\thanks{E-mail: baojiu.li@durham.ac.uk}
\\
$^{1}$Institute for Computational Cosmology, Department of Physics, Durham University, South Road, Durham DH1 3LE, UK
}
\date{Accepted XXX. Received YYY; in original form \today}

\pubyear{2026}
\begin{document}
\label{firstpage}
\pagerange{\pageref{firstpage}--\pageref{lastpage}}
\maketitle

\begin{abstract}
Modern surveys such as DESI and \textit{Euclid}, which collect data for hundreds of millions of galaxies to map the large-scale structure (LSS) of the Universe, hold the key to determining the cosmological parameters and testing new physics. This ambition, however, is limited by uncertainties in the galaxy-halo connection: the link between observed galaxies and the underlying, unobservable matter field, by accounting for effects such as galaxy bias and assembly bias (AB). These are particularly poorly-understood for modified gravity (MG) models, which are popular alternatives to the cosmological constant to explain accelerated expansion. We approach this problem using mock emission line galaxy (ELG) and luminous red galaxy (LRG) catalogues in $f(R)$ gravity matching the specifications of ongoing Stage-IV galaxy surveys, generated from state-of-the-art MG hydrodynamical simulations. While the interplay between MG---especially the chameleon screening mechanism---and galaxy formation leaves complicated imprints in the galaxy-halo connection, a simple physical picture emerges in which halo and galaxy formation are enhanced for progressively more massive haloes over time. We confirm that the basic galaxy-halo connection model, the halo occupation distribution (HOD), in which galaxy occupation is determined solely by halo mass, underestimates galaxy clustering strength in $\Lambda$CDM by $10$--$20\%$ at $z\lesssim1$ when neglecting AB, and demonstrate that MG introduces further complexity. Extending this model with a suitably-chosen environment density as a secondary HOD variable reduces the AB effect in all models to {$2$--$3\%$} for $z\lesssim0.5$. This provides a well-motivated starting point for further works on minimising the impact of AB when testing non-standard cosmological models with LSS.
\end{abstract}

\begin{keywords}
    dark energy -- large-scale structure of Universe -- cosmology: miscellaneous -- cosmology: theory -- galaxies: formation.
\end{keywords}

%%%%%%%%%%%%%%%%%%%%%%%%%%%%%%%%%%%%%%%%%%%%%%%%%%

%%%%%%%%%%%%%%%%% BODY OF PAPER %%%%%%%%%%%%%%%%%%

\section{Introduction}
\label{sec:intro}

The current leading model of cosmology, $\Lambda$CDM, assumes that we live in a flat universe containing cold dark matter (CDM). Furthermore, it assumes that gravity is described by general relativity (GR) with a cosmological constant, $\Lambda$, which drives an accelerating expansion \citep{1999ApJ...517..565P,1998AJ....116.1009R}. Despite the successes of this model, there are several theoretical problems %with the model
that suggest it may not be a complete solution. The lack of a natural physical motivation for $\Lambda$ is among the most well known of these problems. $\Lambda$, with an equation-of-state parameter $w=-1$, has been interpreted as the vacuum energy density. This interpretation wildly clashes with quantum field theory, which suggests a zero-point vacuum energy that is orders of magnitude larger. %Another interpretation is that $\Lambda$ represents a negative pressure fluid, ``dark energy", with 
Modern tests have measured $w$ to high precision. For example, measurements of the cosmic microwave background, such as from COBE, Wilkinson Microwave Anisotropy Probe \citep[WMAP,][]{WMAP2013}, or \cite{Planck15Overview:2016A&A...594A...1P} have shown high levels of agreement with $w=-1$. Observational consistency with $\Lambda$CDM has been further confirmed with measurements of weak-lensing such as from Dark Energy Survey \citep[DES,][]{DESY3CosConstraints:2021arXiv210513546P}, Subaru Hyper Suprime Cam \citep[HSC,][]{HSC_1} and Kilo-Degree Survey \citep[KiDS,][]{KiDS_1}. However, there have been hints that $w$ may deviate from $-1$ implying a time evolution of dark energy \citep[][]{DESI:2025zgx}, and there still remains a tantalising possibility that our description of gravity may be flawed; from this vantage point, a window opens to the world of modified gravity (MG).

There are a plethora of modified theories designed to address a range of theoretical problems, many of which describe alternative expansion histories. We will specifically focus on those intended to serve as alternative explanations for the acceleration of the cosmic expansion, while maintaining a similar background expansion history to $\Lambda$CDM. In other words, we are interested in the remaining freedom of MG to impact large-scale structure (LSS) and galaxy formation for a $\Lambda$CDM-like expansion history. Study of these theories leads to insight into the observational signatures we might expect if the true theory of gravity deviates from the one described by GR with a cosmological constant \citep[see, e.g.,][]{Li2015, Lombriser2014, Hu2013}. In this sense, the main objective is not to treat a specific MG model with the same seriousness as $\Lambda$CDM, but to use some representative examples to illustrate how MG models can be used as a testbed for the validity of our current model of gravity. 

%According to Lovelock's theorem \cite{Lovelock_1}, the Einstein field equations (of GR) are the only local second order equations of motion which can be derived from a 4D action of the metric tensor. This theorem clearly lays out some of the boundaries of what can be achieved with modified gravity, but it also suggests avenues for modifications to GR. If we are to continue to have a theory derivable from an action, then the routes suggested by Lovelock's theorem are: (1) allow higher order terms in the equations of motion \citep[e.g.][e.t.c.]{Lovelock1971, }\Michael{Gauss Bonnet? Lovelock? Weyl? idk if theres any ones interesting for cosmology ($f(R,T)$ or something, but thats more than ones extra degree of freedom often)}, (2) introduce higher dimensions e.g. Dvali Gabadadze Porrati (DGP) braneworld models such as DGP \cite{Dvali2000}, (3) include new fields, e.g. $f(R)$ gravity \citep[e.g][]{Starobinsky1980, Hu:2007PhRvD..76f4004H, }, Chern Simons \citep[][]{Jackiw2003, Smith2008} or Gallileon theories \cite{Nicolis2009}, (4) abandon locality \citep[e.g.][]{Efimov1967, Bahamonde2017}. Inspecting these suggestions, it seems that the most obvious departures from general relativity involve some increase in the complexity of our theory of gravity. 

In this work, we are interested in $f(R)$ gravity, in which we substitute $R$ in the Einstein-Hilbert action for an algebraic function $f(R)$. This can be shown to be equivalent to the introduction of a new scalar field \citep[see, e.g.,][]{Sotiriou:2010RvMP...82..451S}. There are many flavours of $f(R)$ gravity due to the freedom in choosing the function $f$. %, but it is most appropriate to start by choosing functions that only add a single extra degree of freedom to the theory. 
The parameter space of viable choices for $f(R)$ is significantly restricted observationally by tight astrophysical constraints on the solar system scale \citep{Will2014}. %and the speed of gravitational waves from the Laser Interferometer Gravitational-Wave Observatory (LIGO) \citep[e.g.][]{Abbott2017,Abbott2009}. 
We also restrict the parameter space by imposing a requirement for a similar expansion history to that of standard $\Lambda$CDM, though this is already a requirement for screening to take place so that the model can pass Solar system tests \citep[see, e.g.,][]{Brax:2008hh,Wang:2012kj,Ceron-Hurtado:2016jrp}.

LSS is particularly sensitive to the behaviour of gravity across the largest distances. Compared to smaller scales, such as solar system scales, this regime is not well tested. Tests of gravity involving LSS offer constraints that are independent from other astrophysical constraints on gravity, which test different length scales and are subject to different systematics. As such, cosmological tests of gravity, such as those using LSS, work in tandem with astrophysical ones, rather than competing with them. We are now in the era of precision cosmology, with existing data from various observational surveys of galaxies such as Sloan Digital Sky Survey \citep[SDSS,][]{Blanton_2017}, Dark Energy Survey \citep[DES,][]{Abbott2018}, Dark Energy Spectroscopic Instrument \citep[DESI,][]{DESI2016a}, \textit{Euclid} \citep[][]{Euclid:2011arXiv1110.3193L} and the Legacy Survey in Space and Time \citep[LSST,][]{guy2025}. In preparation for the new generation of surveys, %such as DESI \citep{DESI:2016arXiv161100036D} and \textit{Euclid} \citep{Euclid:2011arXiv1110.3193L}, 
it is of vital importance to have precise predictions of galaxy clustering. In the recent DESI data release, measurements of the baryon acoustic oscillations (BAO) seem to indicate a preference for alternative models of dark energy, with $w_0w_a$CDM seeming to be preferred over $\Lambda$CDM \citep{Adame_2025,DESI:2025zgx}. The significance of this has been a subject of debate, with some suggestions that this could be a result of systematics in the supernova data \citep[e.g.,][]{Efstathiou2025}. However, if these deviations are borne out as physical, they would have huge implications for our understanding of the accelerating expansion and could raise interest in MG models that may resolve this problem. 

Dark matter represents the vast majority of the matter content of the universe, and much of %cosmic structure formation 
the LSS can be understood by looking at this component alone. Galaxies form within dark matter haloes \citep{White_78}, which serve as potential wells to capture and condense cold gas and form stars. Furthermore, the physics of dark matter is considerably simpler than that of baryonic matter, which motivates the widespread use of dark matter only (DMO) simulations. Notable examples of DMO simulations in MG include \textsc{elephant} \citep{ELEPHANT_EXAMPLE_1}. The relative computational simplicity of DMO simulations compared to hydrodynamical simulations -- i.e., simulations including baryonic physics -- allows for significantly larger simulated volumes and higher resolutions. This has made them essential for producing LSS predictions of the precision required by the current and upcoming generations of galaxy surveys.

In order to bridge the gap between DMO simulations and observations of galaxies, one of the necessary ingredients is a model of the ``galaxy-halo connection" -- a scheme by which galaxies are ``painted" onto the dark matter haloes of DMO simulations. For an overview of different approaches for modelling the galaxy-halo connection, see \cite{galaxy_halo_review_18}. Using these models facilitates the creation of mock catalogues of galaxies, which serve as an integral piece of pipelines analysing cosmological survey data. One of the most popular methods for creating these mock catalogues is halo occupation distribution (HOD) modelling \citep[e.g.][etc.]{Berlind2003, Zheng_2005}. HOD models assert a direct link between the expected galaxy occupation of a dark matter halo and its mass. These models are computationally inexpensive and flexible, allowing one to essentially ``patch over" the baryonic physics while still enabling constraint of cosmological parameters.

In simple HOD models, the mean galaxy occupancy of a dark matter halo is solely determined by the mass of the halo. In reality, however, there also exist secondary dependencies on other properties of haloes, due to their detailed growth (assembly) history. This latter effect is known as assembly bias (AB). %Assembly bias refers to any dependency of the mean halo occupation on quantities other than the halo mass. Other works 
\cite{Zehavi2018, Hadzhiyska2020} have studied the relationship between halo occupation and secondary parameters, often environment, concentration and formation time. While the HOD models may be flexible enough to ``absorb" these extra dependencies when fitting to observational data, there is no \textit{a priori} guarantee that such an approximate treatment would not bias cosmological parameter estimation. %it remains unclear to what extent they can absorb assembly bias without planting biases in the predictions of cosmological parameters \Michael{I'm now unsure I can say this because the DESI papers did look into this possibility}. This concern should be further considered
This can be particularly true for alternative cosmological models, such as those involving modified gravity, for which there is no detailed understanding of the galaxy-halo connection. If these models exhibit a much stronger assembly bias, standard analysis pipelines may need to be adjusted for tests of MG. Furthermore, using HOD models to create mock catalogues in this way cannot produce accurate predictions on non-linear scales where baryonic effects are of greater importance. 

A more reliable way to understand AB effects in cosmological models is through the use of state-of-the-art hydrodynamical simulations, such as EAGLE \citep{Schaye2014}, IllustrisTNG \citep{Pillepich2017}, and FLAMINGO \citep{Schaye2023}, which are the most physically-grounded approach to modelling galaxy formation, directly following the coupled gravitational and gas dynamics. However, one of the main hurdles for this approach is that the physical processes contributing to galaxy formation occur on a wide range of physical scales, and it is notoriously difficult to deal with large dynamical ranges in numerical simulations. To this end, sub-grid physics models are recipes to summarise the physical processes occurring below resolved scales. These models are essential for realistic modelling of star formation, stellar and black hole feedback, and other processes. Hydrodynamical simulations allow for the robust study of galaxy formation and clustering simultaneously, and therefore they are an ideal candidate for the study of the galaxy halo connection. The validity of the HOD models used for survey mocks can therefore be tested with these simulations. For modern surveys there are analyses performed to quantify the systematic effects of using these models for $\Lambda$CDM \citep[e.g.][ for DESI]{GarciaQuintero2025, MenaFernndez2025}, the validity of these models in the case of modified gravity is not well studied. 

In this work, we attempt to fill this gap by studying the galaxy-halo connection and assembly bias effect on galaxy clustering in a wide range of $f(R)$ models with varying degrees of deviation from $\Lambda$CDM, for different galaxy types and a range of redshifts. Our aim is to (i) understand qualitatively how MG affects galaxy formation using high-resolution cosmological galaxy formation simulations, (ii) quantify the level of AB effect and check its gravity dependence, and (iii) find ways to reduce the AB effect in clustering predictions, by extensions to the simple HOD model.

This paper is organised as follows. In Section \ref{sect:model_sim} we briefly introduce $f(R)$ gravity and the hydrodynamical simulations used in this work. In Section \ref{sect:catalog_generation} we describe the galaxy catalogues to be used, luminous red galaxies and emission line galaxies, and how they are generated from the simulations. We then describe the HOD model along with its key assumptions in Section \ref{sect:hod_modelling}, before proceeding to a detailed inspection and analysis of the redshift and gravity dependencies of the HOD parameters in Section \ref{sec:HOD_fits}. In Sections \ref{sec:HOD environment effect} and \ref{sec:HOD concentration effect} we offer a detailed breakdown of the HOD behaviour in bins of secondary halo properties, halo environmental density and concentration, respectively, to understand how these additional halo properties affect galaxy formation in different gravity models. Based on these splits into bins of secondary halo properties, we extend the simple HOD model to include these properties, and assess their impact on the assembly bias effect in galaxy clustering, in Section \ref{sect:ab}. We find that the HOD extension with halo environment density as a secondary HOD variable can eliminate a majority of the AB effect. Finally, we discuss and conclude in Section \ref{sec:conc_and_disc}.

\section[MG models and simulations]{Modified gravity theory and simulations}
\label{sect:model_sim}

\subsection{$f(R)$ gravity}
\label{subsect:fR_model}

The $f(R)$ model is obtained by a substitution of the Ricci scalar, $R \to R+f(R)$, in the Einstein-Hilbert action:
\begin{equation}
S_{f(R)} = \int \sqrt{-g}\left[\frac{1}{16\pi G}\left(R +f(R)\right) + \mathcal{L}_m\right]\text{d}^4x,
\end{equation}
where $g$ is the determinant of the metric tensor, $g_{\mu\nu}$, $\mathcal{L}_m$ is the matter Lagrangian density, and $G$ is Newton's constant. 

There are two versions of $f(R)$ gravity, depending on whether the field equations are derived by minimising the above action with respect to the metric tensor only, or both the metric and the connection coefficients, $\Gamma^\mu_{\nu\rho}$. These are respectively known as the metric and the Palatini formulation of $f(R)$ gravity. For $f(R)=\text{const.}$, the two formulations are equivalent to each other, however in general they are different theories of gravity. Here, we are interested in the metric formulation which has the following Einstein field equation:
\begin{equation}
    \label{eqn:EFE}
    G_{\mu\nu} + f_R R_{\mu\nu} - \left[\frac{1}{2} f(R) - \Box f_{R}\right]g_{\mu\nu} - \nabla_\mu \nabla_\nu f_{R} = 8\pi G T^m_{\mu\nu},
\end{equation}
where $\nabla_\mu$ is the covariant derivative, and $\Box\equiv\nabla^\mu\nabla_\mu$, $R_{\mu\nu}$ is the Ricci tensor, $G_{\mu\nu} \equiv R_{\mu\nu} - \frac{1}{2}Rg_{\mu\nu}$ is the Einstein tensor, and $T^m_{\mu\nu}$ is the matter energy-momentum tensor. 
The equation of motion for the new scalar field, $f_R$, can be obtained by taking the trace of Eq.~\eqref{eqn:EFE}:
\begin{equation}
    \label{eq:EOM fR}
    \Box f_R = \frac{1}{3}(R - f_R R + 2f(R) + 8\pi G \rho_m),
\end{equation}
where $\rho_m$ is the density of matter. This equation can be interpreted as the equation of motion of a new scalar degree of freedom, a scalaron, $f_R$. 

We will work within the quasi-static approximation, which assumes time derivatives of $f_R$ are small and can be neglected \citep{Bose_2015}. For the models we are working with, $|f_R| \ll 1$ and the effect on the background cosmology is practically indistinguishable from that of a cosmological constant $\Lambda$. We work in the Newtonian gauge, giving the following equations:
\begin{equation}
    \label{eq:EOM Phi}
    \vec{\nabla}^2 \Phi \approx \frac{16\pi G}{3} \delta \rho_m a^2 + \frac{1}{6} \delta R,
\end{equation}
\begin{equation}
    \label{eq:f_R_poisson}
    \vec{\nabla}^2 f_R \approx - \frac{1}{3} \left[ \delta R +8\pi G \delta \rho_m a^2\right],
\end{equation}
where $\vec{\nabla}$ is the spatial gradient, $\Phi$ is the Newtonian potential, $a$ the scale factor, and $\delta Q$ means the perturbation around the background value, $\bar{Q}$, of a quantity $Q$. 

These two equations can be combined to get a different form of the modified Poisson equation:
\begin{equation}
    \vec{\nabla}^2 \Phi \approx {4\pi G}\delta \rho_m a^2 + \frac{1}{2}\vec{\nabla}^2f_R,
\end{equation}
which highlights that $\frac{1}{2}f_R$ can be considered as the potential of a ``fifth force''.
%Eq.~\eqref{eq:f_R_poisson} represents an $f(R)$ equivalent to the Poisson equation, which describes Newtonian gravitation, which GR reduces to. However, here the constant multiplying $\rho_m$ is a factor $4/3$ larger. The concept of "screening" in modified gravity is an important theoretical consideration. 
There must be potentially measurable effects of this fifth force in particular situations for a theory to be interesting. However, in well-studied regimes such as the Solar system -- where GR has been shown to agree very well with experiments --  there must be no such fifth-force effects. It is therefore common for MG theories to feature a mechanism by which the equations of motion reduce to those of GR in environments such as the Solar system. 

$f(R)$ gravity is part of a family of gravity models exhibiting the so-called chameleon screening mechanism \citep{Khoury:2004PhRvD..69d4026K,Khoury:2004PhRvL..93q1104K}. In low-density regions, $|\delta R| \ll 32\pi G |\delta \rho_m|a^2$, Eq.~\eqref{eq:EOM Phi} reduces to $\vec{\nabla}^2 \Phi \approx \frac{16}\pi G\delta \rho_m a^2/3$, implying a factor-of-$1/3$ enhancement of gravity. In sufficiently high-density regions, 
the scalar field $f_R$ settles to $0$, and according to Eq.~\eqref{eq:f_R_poisson} $\delta R \approx - 8\pi G \delta \rho_m a^2$ which, if put in Eq.~\eqref{eq:EOM Phi}, recovers the GR case $\vec{\nabla}^2 \Phi \approx 4\pi G\delta \rho_m a^2$. As a result, very massive objects, and objects in high-density environments, exhibit screening, whereas smaller objects and objects in low-density environments experience a modified gravity force that is up to $1/3$ stronger than in GR. This suggests that gravitational processes, such as those involved in galaxy formation or generally in the formation of LSS, can be impacted by MG. Due to the strong environmental dependence, such impacts can be accurately assessed only through numerical simulations.

Here we work with Hu-Sawicki $f(R)$ gravity, which is defined by an $f(R)$ of a broken power law form \citep{Hu:2007PhRvD..76f4004H}:
\begin{equation}
    \label{eq:Hu-Sawicki}
    f(R) = -m^2 \frac{c_1 (R/m^2)^n}{c_2(R/m^2)^n + 1},
\end{equation}
with $m^2 \equiv H_0^2 \Omega_{\textrm{M}}$ and $\Omega_{\rm M}$, $H_0$ being the matter density parameter and the Hubble constant today, and $n, c_1$ and $c_2$ are free parameters. We take $n=1$ for illustration purposes.

In the regime where $|R| \gg m^2$, which is relevant on cosmological scales, the function $f(R)$ has a nearly constant value. In order to match the background cosmology of $\Lambda$CDM, %and furthermore maintain agreement with precision measurements of cosmology, 
we set $\frac{c_1}{2c_2}m^2 \approx \Lambda$, leaving only one model free parameter. From Eq.~\eqref{eq:Hu-Sawicki}, the scalar field $f_R$ can be written as:
\begin{equation}
    f_R =\frac{\dd f(R)}{\dd R} = - n \frac{c_1}{c_2^2}\left( \frac{m^2}{R} \right)^{n+1}.
\end{equation}
Following the convention in the literature, we take the one free parameter to be $|f_{R0}|$, which is the magnitude of the background value of the scalar field today. It can be shown that
\begin{equation}
    \left|f_{R0}\right|  = \frac{c_1}{c_2^2}\left(\frac{\Omega_{\textrm{M}}}{\Omega_{\textrm{M}}a^{-3}+4\Omega_{\Lambda}}\right)^2,
\end{equation}
where $\Omega_\Lambda=1-\Omega_{\rm M}$. We conform to the literature nomenclature and refer to our models as `F$x$', corresponding to $|f_{R0}| = 10^{-x}$. Models with larger $|f_{R0}|$ experience a stronger fifth-force effect because their screening is less efficient (though the fully unscreened fifth force always has $1/3$ the strength of Newtonian gravity for any $f_{R0}$ value), and as such we refer to them as ``stronger" models. Models with smaller $|f_{R0}|$ are ``weaker'' models.

\subsection{Galaxy formation simulations in $f(R)$ gravity}
\label{subsect:galaxy_sims}

The environmental dependence of $f(R)$ models makes their cosmological behaviour highly nonlinear. Mathematically, this is reflected by the strong nonlinearity of Eq.~\eqref{eq:EOM fR}. This means that an $f(R)$ simulation is much more expensive than an $\Lambda$CDM simulation with the same specifications. While N-body simulations in chameleon-type theories first appeared nearly two decades ago \citep{Oyaizu:2008sr,Li:2009sy,Zhao:2010qy,2012JCAP...01..051L}, simulations with fully calibrated galaxy formation physics have appeared much later \citep[e.g.,][]{2019NatAs...3..945A,Hernandez-Aguayo:2020kgq}. \cite{Mitchell_2022} developed the first suite of full-physics simulations in $f(R)$ gravity in a cosmological volume, allowing to reliably study large-scale clustering of galaxies and various statistics of galaxy clusters in this model.

We have later extended the simulations of \cite{Mitchell_2022} by running further variations of the $f_{R0}$ parameter --- we now have values $\log_{10}(\mid f_{R0} \mid) = -6.0, -5.5, -5.0, -4.5, -4.0$ which we refer to as F6.0, F5.5, F5.0, F4.5, F4.0 respectively, following the above convention. This simulation suite also has $\Lambda$CDM, which is equivalent to $f_{R0} = 0$. While some of these models may be incompatible with some current astrophysical constraints \citep[e.g.,][]{Baker:2019gxo}, cosmological and astrophysical constraints are complementary as they involve different time and distance scales. Furthermore, they are subject to different systematic effects as well as theoretical uncertainties. Since our objective is mainly to adopt $f(R)$ gravity as a concrete example to demonstrate the complicated effect of MG on galaxy formation, the above parameter choice is advantageous since it covers a variety of chameleon screening strengths. We will see the rich phenomenology these parameter choices result in below.

The suite of simulations were carried out using a modified version of the \textsc{arepo} code \citep{2010MNRAS.401..791S}, \textsc{mg-arepo} \citep[see][for a detailed description]{2019NatAs...3..945A}, which includes a modified gravity solver. They also feature a retuning of the IllustrisTNG full baryonic physics model \cite{Pillepich2017}. The subgrid physics model has been recalibrated for the resolution of $1136^3$ dark matter particles in a $301.75\,h^{-1}\mathrm{Mpc}$ box, which is equivalent to a dark matter particle mass of $1.35 \times 10^9 \, \mathrm{M_\odot}$. This resolution is lower than that of IllustrisTNG (specifically, TNG300-1), but broad statistical agreement is achieved with 6 observables: the stellar mass function, stellar-to-halo mass relation, cosmic star formation rate density, cluster gas mass fraction, and correlation between galaxy size and black hole mass with stellar mass \citep[see Fig. 2 and Fig. 3 of][]{Mitchell_2022}. Note that all models used the same subgrid physics: it was found that there was no need to recalibrate the physics model for different values of $f_{R0}$, since the impact of MG on the calibrated subgrid parameters is much smaller than the observational uncertainties in these observables. For details about the subgrid physics, readers are referred to \cite{Mitchell_2022}. These simulations were previously used to analyse the impact of MG on large-scale galaxy clustering \citep[][hereafter Paper I]{Collier_2024}.

We use the Planck best-fit cosmological parameters in all runs $(h,\Omega_{\rm M},\Omega_{\rm B},n_{\rm s},\sigma_8) = (0.6774, 0.3089,0.0486,0.9667,0.8159)$, where $h=H_0/\left(100\,\text{km/s/Mpc}\right)$ is the dimensionless Hubble constant, $\Omega_{\rm B}$ is the density parameter of baryons in the present day, $n_{\rm s}$ is the primordial density power spectrum's spectral index and $\sigma_8$ is the root-mean-square of the fluctuations of the matter density field today, smoothed on a scale of $8h^{-1}\textrm{Mpc}$.

As in \citet{Mitchell_2022}, halo catalogues used across this work were identified using the \textsc{subfind} code \citep[][]{springel2001populating}, which is part of the \textsc{arepo} package. This code uses the friends-of-friends (FoF) algorithm to find FoF groups (haloes) and then uses a gravitational unbinding method to identify the bound substructures (i.e., the subhaloes) in each group. Galaxies correspond to bound stellar mass contained within these subhaloes.

\section{Galaxy Catalogue Generation} \label{sect:catalog_generation}

We generate catalogues for two different populations of galaxies -- luminous red galaxies (LRGs) and emission line galaxies (ELGs) -- at $z \lesssim 1.2$, comparable to the redshift range for these populations in current cosmological surveys. An extended redshift range will help us to better explain and justify differences between gravity models, as we shall see below.

\subsection{The LRG catalogues}
\label{subsect:lrg_cat}

LRGs are usually quiescent and passively evolving galaxies, highly massive and photometrically red in colour. They are a primary target of the DESI spectroscopic survey, particularly at low redshift. 

Our LRG catalogues are created by ranking simulated galaxies by stellar mass and taking the top $N = n_{\rm g} \times L^3$ galaxies, where $n_{\rm g}$ is our target number density, and $L$ is the box size of the simulation along each dimension. The stellar mass is defined to be the summed mass of all star particles within $2\times r_{1/2 \star}$, where $r_{1/2 \star}$ is the radius containing half of the stellar mass of the corresponding subhalo.

\subsection{The ELG catalogues}
\label{subsect:elg_cat}

ELGs, so-named for the presence of strong emission lines in their spectra, are typically lower mass, actively star-forming galaxies, and are another primary target of the DESI spectroscopic survey.

For ELGs, we use the simplification presented in \cite{Hadzhiyska2021}: we identify an ELG-like population by using a sample selected on specific star formation rate (sSFR), rather than one selected by cuts in colour and magnitude, as is formally done within surveys like DESI. The cuts adopted for ELG populations in cosmological surveys are generally designed to select high sSFR objects, and so this approach is well motivated. \cite{Hadzhiyska2021} also show that the resulting population has similar HOD and clustering properties within the redshift range of these surveys. 

At lower redshifts there is a more notable discrepancy between the properties of samples generated via sSFR cuts and those generated via colour cuts. Determining the colours of galaxies requires the use of stellar population synthesis models, such as flexible stellar population synthesis \citep[see, e.g.,][]{Conroy2010}, and some modelling of resolved dust. Due to the low resolution of the simulations used in this work, it is possible that the star formation histories of the lowest-mass galaxies will be unreliable, particularly at early epochs when they are less well resolved. For this reason, it is necessary to apply a stellar mass cut to this sample when generating ELG samples using this method. In \cite{Hadzhiyska2021}, a mass cut of $~ 1.0\times10^{10}\,\mathrm{M_\odot}$ was imposed, which was calibrated by comparison with colour-selected samples; the exact cut varied with redshift slightly. In Paper I \citep[][]{Collier_2024}, we used a higher mass cut of $M_\star\sim2.5\times10^{10}\,\mathrm{M_\odot}$, which excludes poorly-resolved objects with fewer than $\~100$ star particles.

\begin{figure*}
    \centering
    \includegraphics[width=\linewidth]{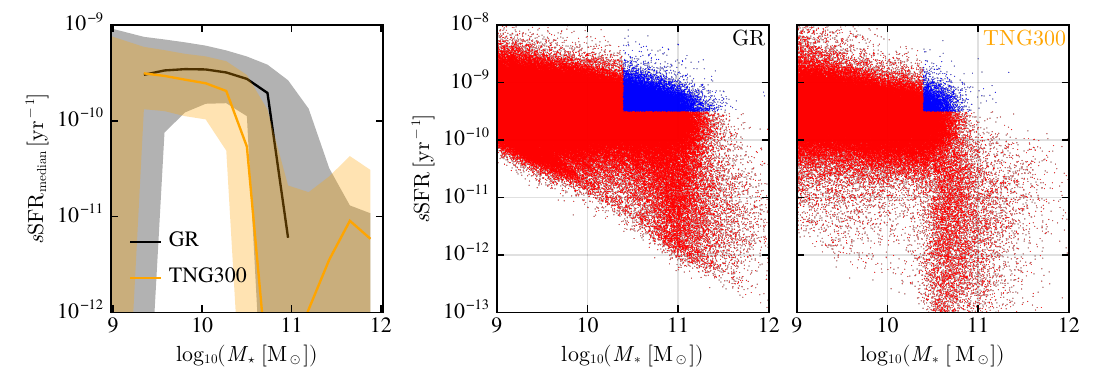}
    \caption{The leftmost panel shows the median sSFR in bins of stellar mass for our GR simulation and for TNG300, where the shaded regions show the $16\%$th and $84\%$th percentiles. The central and rightmost panels show the sSFR and stellar masses plotted for subhaloes in both our GR run and TNG300. Blue points show objects corresponding to cuts $M_\star > 3 \times 10^{11} \, \mathrm{M}_\odot$ and $s\mathrm{SFR}> 10^{-9.5} \, \mathrm{yr}^{-1}$. GR and TNG models are both shown at $z\approx 0.8$: $z=0.80$ and $z=0.81$ for GR and TNG, respectively.}
    \label{fig:sSFR_vs_stellar_mass}
\end{figure*}

As a first check of our galaxy-selection method, we compare the star-forming properties of subhaloes in our GR run to those of the TNG300 simulation \citep[][]{Nelson:2018uso}---comparing the distribution of sSFR vs stellar mass for the subhaloes in each simulation---in Figure~\ref{fig:sSFR_vs_stellar_mass}. For clarity, we stress that TNG300 should not be considered the ``truth''---nevertheless, the higher resolution of the TNG300 simulation serves as a useful baseline against which to compare the star-formation properties of our ELG sample, and to determine the mass scale below which resolution effects become important.

The lower resolution of our simulations leads to a cutoff sSFR below which there are no objects, which manifests as a ``stripe" in the low-sSFR regime that is caused by objects with just a single star-forming gas cell. Above this stripe, the shape of the sSFR-$\log_{10}M_\star$ distribution is visibly similar to that of TNG300, but notably there seems to be a greater prominence of high stellar mass, star forming subhaloes in our simulation. This 
can be seen clearly in the blue-coloured points and in the left panel of Fig.~\ref{fig:sSFR_vs_stellar_mass}, and is likely because of the recalibration of our subgrid physics model to the lower resolution. 

We have stayed well away from the ``stripe" at low sSFR, where we expect to have much larger shot noise on the star formation. We find, by using a rudimentary check at $z = 0.8$, that making a stellar mass cut of $2.5\times 10^{10} \, \mathrm{M_\odot}$ is sufficient to make samples of number density $0.001 \, h^3\mathrm{Mpc}^{-3}$ with $\log_{10} (\text{sSFR}\, \mathrm{[M_\odot yr^{-1}]} ) >-9.5$ in both TNG300 and our simulation. This should allow to reasonably select ELGs. 

Our simulation has a greater abundance of star-forming objects even after removing galaxies with low star formation. This is likely a consequence of the presence of star-forming subhaloes at higher mass than in TNG300. We compared the clustering for TNG300 and our simulation for the ELG samples defined this way, and found that on large scales both are in reasonable agreement (well within $5\%$ on scales relevant in this work, though this figure is purely indicative and should not be interpreted as the ``error'' of our simulations since there is no ``ground truth'' here.). This suggests that despite the differences observed in Fig. ~\ref{fig:sSFR_vs_stellar_mass} and the differences in resolution, the galaxy populations themselves behave similarly in both simulations. Disagreements in small-scale clustering are an expected effect when comparing simulations with different resolutions. We found that the agreement between TNG300 and our simulation for ELG clustering is independent of the cuts.

\begin{table}
    \begin{center}
    \begin{tabular}{ |c|c|c| } 
     \hline
      & ELG & LRG \\ 
     \hline
     $z$ & $\log_{10}(\mathrm{sSFR}\, [\rm yr^{-1}])$ & $M_{\star} \, [\rm 10^{10}\, M_\odot]$ \\ 
     \hline
     $1.16$ & $-9.3$ & $7.5$ \\
     $0.97$ & $-9.4$ & $7.4$ \\
     $0.73$ & $-9.5$ & $8.2$ \\
     $0.51$ & $-9.6$ & $8.2$ \\ 
     $0.27$ & $-9.9$ & $8.6$ \\ 
     \hline
    \end{tabular}
    \end{center}
    \caption{Sample cuts for LRG and ELG samples for a selection of our catalogs. These cuts are for our $\Lambda$CDM samples. We note that ELG samples contain a constant stellar mass cut of $M_\ast = 2.5\times10^{10} \, \rm M_\odot$.}
    \label{tab:example_cuts_table}
\end{table}
We present a table of the sample cuts for a selection of our populations from the GR runs in Table~\ref{tab:example_cuts_table}. The cuts vary with redshift due to our enforced requirement of constant number density.

\section{HOD modelling}
\label{sect:hod_modelling}

\subsection{An outline of the basic HOD model}

The halo occupation distribution (HOD) model is a treatment of the galaxy-halo connection, popular for its speed, simplicity and flexibility. At its core, it assumes that the occupancy number---the count of galaxies within a dark matter halo---depends solely on the mass of that halo. There are three ingredients that define an HOD model: 
\begin{enumerate}
    \item the distribution mean as a function of halo mass; \label{distpoint1}
    \item the distribution shape; and \label{distpoint2}
    \item the radial distribution of objects within the halo. \label{distpoint3}
\end{enumerate}
HOD models generally split modelling of central and satellite galaxies. Centrals for our purposes will refer to the galaxy hosted by the subhalo in the FoF group with the largest total mass, while galaxies in the remaining subhaloes are categorised as satellites. For clarity, it is possible (and in some cases expected) to have a halo with satellite galaxies but no central galaxy. Regarding point \ref{distpoint1}, in cosmological survey analysis the use of these models often involves a choice of a functional form of the HOD, and then, one way or another, the fit parameters are optimised in order to best fit measured clustering. The resulting HOD fit parameter values are generally categorised as ``nuisance parameters", and considered to be physically uninteresting. These parameters will ultimately arise from some combination of physical and selection effects, but it will likely do so in a complicated way that may not have a straightforward physical intuition.

In Section \ref{sect:catalog_generation} we described the two galaxy populations of interest here, ELGs and LRGs, for which different HOD models are prescribed. %We will now describe our chosen models of interest. 
The simpler case is LRGs, for which we use the 5 parameter model of \cite{Zheng_2005}:
\begin{equation}
    \label{eq:basic_hod_central}
    \langle N_{\rm cen}\rangle = \frac{1}{2}\left[1+\mathrm{erf}\left(\frac{\log M_h - \log M_{c}}{\sigma_{\log M}}\right)\right],
\end{equation}
\begin{equation}
    \label{eq:satellite_HOD_model}
    \langle N_{\rm sat}\rangle = \left( \frac{M_h-M_{\rm cut}}{M_1}\right)^\alpha,
\end{equation}
which describes, respectively, the mean central and satellite occupancy. Here $\textrm{erf}(x)$ is the error function.

%The mean central occupation changes from $0$ to $1$ around $M_{c}$ across a characteristic width $\sigma_{\log M}$. This can be conceptually understood as in larger haloes AGN feedback heats the ISM and quenches star formation, so that we expect virtually all galaxies above a certain halo mass to be red. Galaxies in smaller haloes have lower stellar mass, and are thus too faint to meet magnitude cuts in surveys; as such the HOD is 0 for smaller halo masses. Satellite LRG occupations are modelled with a simple power law with index $\alpha$ above a cutoff mass $M_{\rm cut}$, with haloes having zero occupancy below that cutoff.

\begin{figure*}
    \centering
    \includegraphics[width=\linewidth]{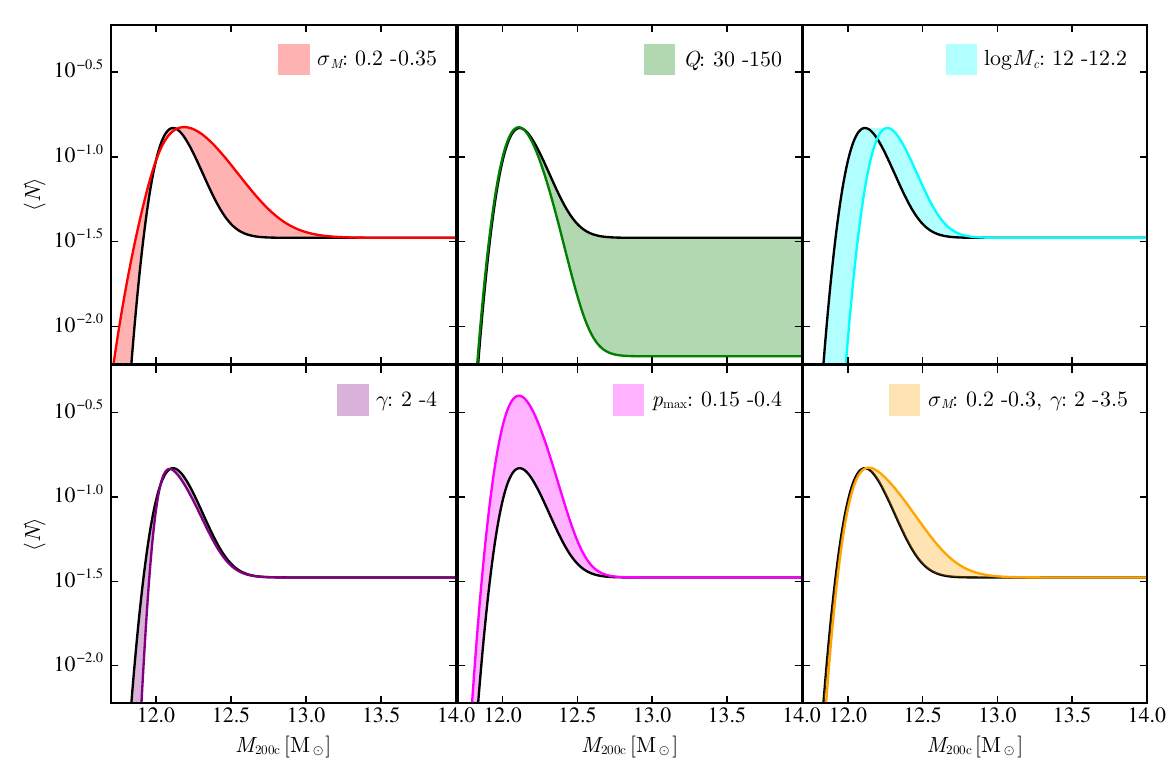}
    \caption{A demonstration of the effect of varying the different HOD parameters on the central ELG mean occupancy. The black curve corresponds to parameter values $(\sigma_M,\gamma,Q,p_{\rm max},\log M_c) = (0.2, 2.0, 30.0, 0.11, 12.0)$, i.e., the lowest value in the range of variation of each parameter in this figure. The shaded regions show the space swept out by gradually increasing the value of each parameter over the range shown in the top right legend, with solid coloured curves designating the curve associated with the maximum value of the parameter.}
    \label{fig:parameter_vary_demo}
\end{figure*}

For ELGs, we use the high mass quenching (HMQ) model of \cite{alam_2020}:
\begin{equation}
    \label{eq:HMQ_HOD_central}
    \begin{aligned}
    &\langle N_{\rm cen} \rangle = 2A\phi(M_h)\Phi(\gamma M_h) + \frac{1}{2Q}\left[1 + \mathrm{erf}\left(\frac{\log M_h - \log M_c}{0.01}\right)\right],\\
    &\mathrm{where}\\
    &\phi(x) = \mathcal{N}(M_c,\sigma_M),\\
    &\Phi(x) = \int_{-\infty}^x\phi(t) \dd t = \frac{1}{2}\left( 1+\mathrm{erf}\left[ \frac{x}{\sqrt{2}}\right]\right),\\
    &A = \frac{p_{\rm  max} - 1/Q}{\max(2\phi(x)\Phi(\gamma x))}.
    \end{aligned}
\end{equation}
The HMQ model uses the same prescription for satellites as for LRG,  Eq.~\eqref{eq:satellite_HOD_model}. %, for the modelling of ELG satellite occupations. 
The first term in Eq.~\eqref{eq:HMQ_HOD_central} is a skewed Gaussian curve centred about $M_c$ with width  $\sigma_M$, with the skewness controlled by the $\gamma$ parameter. The second term characterises the plateau in occupations at high masses, with a plateau height of $1/Q$. Lastly, $p_{\rm max}$ determines the height of the peak in the central occupation. This is an important region of the HOD since the HMF increases as a power law as the halo mass decreases. Due to the complexity of this model, we provide a demonstration of the influence of each fit parameter in Fig.~\ref{fig:parameter_vary_demo}. 

Now we speak about point \ref{distpoint2}. The nature of the distribution is generally chosen for simplicity. For example, in a basic HOD model it is common to choose the central occupation as a single Binomial distribution, and satellites with a Poisson \citep[][etc]{Avila2018, Zehavi2005, Smith2017}. One benefit of these choices is that both distributions are fully specified by their means, and so no extra parameters are introduced to the model.

Finally, for point \ref{distpoint3}, there are a few schemes used to place the galaxies in haloes. One popular approach is to use the measured NFW density profile of the halo as a distribution. This requires some truncation, as the integral of the NFW profile does not converge over all space (although in practice this is not a problem as haloes do not actually have an infinite extent). Another popular scheme is to grab a random dark matter particle associated with the halo, and place a galaxy at its position. This samples the true dark matter distribution, as opposed to the fitted NFW profile. 

\subsection{Assumptions of the basic HOD model}\label{subsec:assumptions_basic_HOD}
The basic HOD model outlined in \ref{distpoint1}-\ref{distpoint3} contains several assumptions, both implicit and explicit. In this subsection we will outline these assumptions and their validity. 
\subsubsection{Assumption 1: Occupations depend only on halo mass}

This is a simplification that greatly contributes to the flexibility of basic HOD models, but also an approximation that is known to be inaccurate. Other works often experiment with extra parameters into the HOD model, such as halo formation time, halo environment, and halo concentration. It is worth noting that the dependency of halo occupations on extra parameters does not in itself imply assembly bias. Ignoring the dependency of the HOD on such additional parameters is therefore not inherently problematic. 

Furthermore, even if there is assembly bias, it is not necessarily a problem so long as it does not bias ``non-nuisance" parameters. However, it is possible that, due to degeneracies between HOD and cosmological parameters, one may infer incorrect values for the latter. For this reason, if there is hint for AB effects, it is worth investigating how this might impact the inferred cosmological parameters. Alternatively, as mentioned above, one can investigate the HOD dependencies on secondary parameters to try to develop improved HOD models that account for the assembly bias. 

\subsubsection{Assumption 2: Poissonian distribution of satellites}\label{subsubsec:poissonian assumption}

Poissonian distributions have the property that $\langle N \rangle = \mathrm{Var}[N]$, so that one way we can characterise deviations from Poissonian occupations is to look at the ratio between the variance and the mean of the true distribution: if $\mathrm{Var}[N]/\langle N \rangle < 1$ ($>1$) this is referred to as a sub (super)-Poissonian distribution. It was found in \cite{Hadzhiyska2023} that ELGs display super-Poissonian statistics, and in this paper we find a similar result. The consequence of this for clustering is primarily for smaller scales --- high occupancy haloes disproportionately contribute to pair counts in the one-halo part of the correlation function. In this work, we use the measured true distribution of satellite occupations in order to avoid this problem.

\subsubsection{Assumption 3: Satellite and central occupations are independent (one-halo conformity)}
\label{subsubsec:cen-sat independence}

It is known that galaxies form cooperatively \citep{Bower1993}: the formation of one galaxy is thought to enhance the formation of nearby galaxies. Hence, for blue, star-forming, galaxies, one expects that the presence of one central ELG will enhance the probability of finding a satellite ELG. This is another factor that would primarily matter for smaller-scale clustering, however. This effect will be ignored in this work.

\subsubsection{Assumption 4: Satellite positions correspond to the dark matter distribution of the halo}\label{subsubsec:satellite radial distribution}

As mentioned above, two common methods of placing satellites in haloes involve either sampling the fitted NFW profile of the halo, or directly sampling the true dark matter distribution by placing galaxies at the positions of random particles of the halo. We find that the distribution of satellite galaxies does not quite conform to the dark matter distribution, particularly in the case of ELGs. ELG satellites are expected to preferentially exist on the outskirts of their host halo, as they are less likely to be quenched early in the infall. To account for this effect, we will use the true radial satellite distribution measured directly from the hydrodynamical simulations.

\subsubsection{Assumption 5: Satellite positions are not correlated}

Due to the nature of cooperative galaxy formation mentioned above,  one would also expect ELG satellites to typically exist close to one another. This is a smaller effect compared to some of the other assumptions here, so we have decided not to account for this.

\section{Fitting HODs} 
\label{sec:HOD_fits}

\subsection{The HOD and its general behaviour}
\label{subsect:general_hod_trend}

We measure the mean occupation of haloes and its standard error from simulations as follows:
\begin{itemize}
    \item From the galaxy catalogues, we identify the host haloes of the galaxies, from which we compute the central and satellite galaxy occupancy of each halo.
    \item For each halo, we compute the dark matter mass within $R_{\rm 200c}$, i.e. the radius from the halo centre of mass within which the average density is $200\times\rho_{c}$, where $\rho_c$ is the critical density of the Universe. We will denote mass as $M_{\rm 200c}$ in line with common nomenclature. We then compute the mean satellite and central occupancies of haloes within bins of $M_{\rm 200c}$.
    \item We finally perform maximum likelihood fitting on the measured HODs, using the previously-mentioned HOD models. We assume Poisson distributed satellite occupation and Bernoulli distributed central occupation for our likelihood functions in fitting. While satellite occupations are not truly Poisson distributed, we deem this acceptable for the purposes of fitting\footnote{Note that a Poissonian distribution is assumed only in the HOD parameter fitting. When generating mock galaxy catalogues using the HOD model, we use the actual measurement of the satellite profiles, as mentioned above.}. 
\end{itemize}

%I dont actually think this next paragraph is necessary by any means, but pieces of it may be salvageable for other parts, so I leave it commented!
%One of the main uses of HOD models is the generation of mock galaxy catalogs. The models are imposed onto halo catalogs from DMO simulations in order to produce much larger mock galaxy catalogs than would be possible with hydrodynamical simulations, which are much more computationally expensive. One obstacle of this approach is that there is some feedback from the baryonic physics onto the dark matter, which results in some variation in halo properties (e.g. the HMF) between DMO and hydrodynamical simulations. The problem is commonly remedied by applying some mass correction procedure to the DMO haloes. We note that the since the HODs measured here are directly from a hydrodynamical simulation, we do not need to include any such correction.

\begin{figure*}
    \centering
    \includegraphics[width=\linewidth]{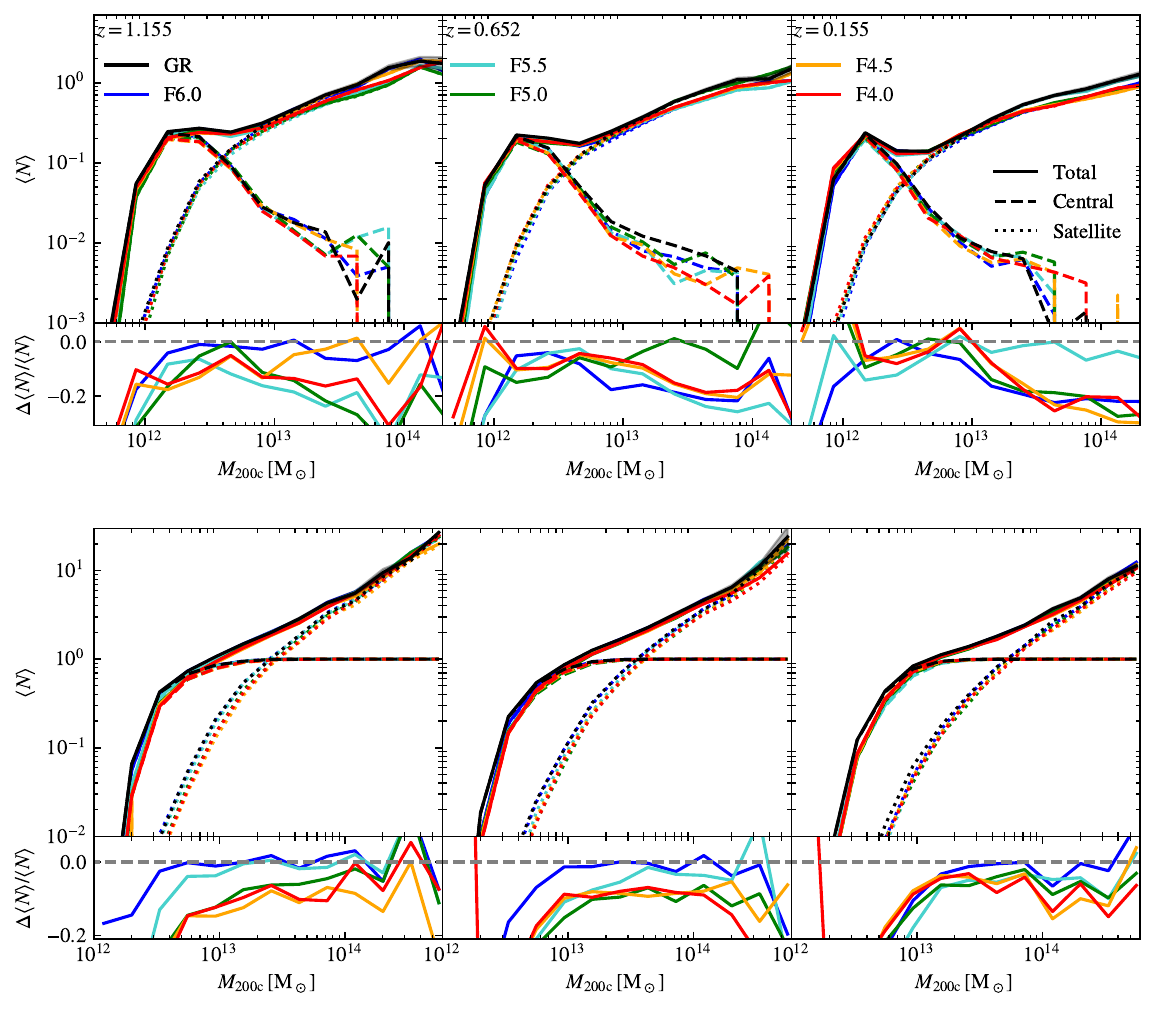}
    \caption{ 
    \textit{Upper row:} Mean occupations of ELG-hosting haloes as a function of $M_{\rm 200,c}$. Different line styles show the total, central and satellite occupations as indicated by the legend in the rightmost subpanel. The shaded region shows the standard error on the mean total occupation for GR, inflated by a factor of $5$ to aid visibility. The smaller subpanels show the relative difference between each MG model and GR. \textit{Lower row:} The same but for the LRG halo occupation. In all cases the galaxy number density is $n_g = 0.001\, h^3\rm Mpc^{-3}$, with decreasing redshift from left to right as indicated at the top-left corner of each ELG top subpanel. The line colours represent the gravity model; the same colour scheme will be used throughout this paper.} 
    \label{fig:HOD examples}
\end{figure*}

Figure \ref{fig:HOD examples} shows a few examples of the HODs measured at different redshifts for each of the models. First of all, and this holds true for both LRGs and ELGs, the different gravity models all have similar enough general shapes to GR that we expect the same functions to be appropriate for fitting the MG models, albeit with differences in the values of the fitted HOD parameters. Furthermore, we note that the MG model HODs are almost always lower than in GR. We will try to give physical interpretation to the results below.

As we shall see repeatedly below, the effect of MG on the HOD does not always follow a monotonic pattern, with the strongest deviation happening for F4.0. However, neither is it completely random. Consider the LRG HODs first, shown in the bottom row of Fig.~\ref{fig:HOD examples}. Here the MG effect is actually monotonic, and can be explained by the enhanced HMF. In modified gravity theories, halo abundance is larger around the relevant mass scale -- see, e.g., Figure 3 of Paper I. This makes the mean occupancy smaller. We also note the ``saturation'' effect described in Paper I, where the MG effect seems to have an upper limit, e.g., in the three snapshots F4.5, F5.0 and F5.5 respectively catches up with F4.0.

The behaviour of the ELG HODs is more complicated and on first sight seems random. Upon closer inspection, however, it can be noted that the deviation from GR follows an ``evolution sequence''. We will discuss this in more detail in the next subsection when plotting the time evolution of the fitted HOD parameters. 

\subsection{Time evolution of ELG HOD}
\label{subsect:elg_hod_evolution}

\begin{figure*}
    \centering
    \includegraphics[width=\linewidth]{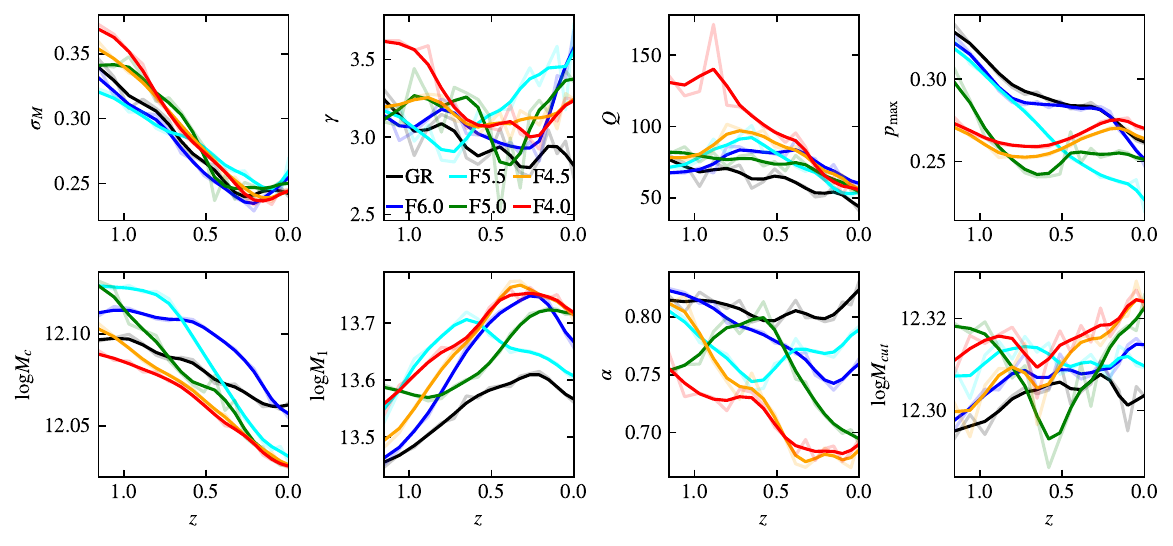}
    \caption{Best-fit ELG HOD parameters as a function of redshift. The HODs correspond to galaxy populations with a fixed number density of $0.001 \, h^{3}\mathrm{Mpc}^{-3}$. The solid curves show the best-fit parameters smoothed over redshift using Gaussian smoothing with $\sigma = 1.5 \,\rm$ plotted points. % \Sownak{what do you mean by 1.5 `points'?}\Michael{Essentially it uses a gaussian kernel on a grid with our data points evenly spaced, and it weights them according to this kernel with characteristic sigma = 1.5 of those grid points. Does "plotted points" make this clear enough?}. 
    The unsmoothed curves are shown underneath as faded curves of the same colour.}
    \label{fig:ELG_fit_params}
\end{figure*}

In Figure ~\ref{fig:ELG_fit_params} we show the fitted ELG HOD parameters for all models as functions of redshift. We refer readers to Fig. ~\ref{fig:parameter_vary_demo} for a demonstration of the impact of the ELG HOD parameters on the shape of the occupancy. In trying to provide physical explanations, we will strive to focus on the parameters which are most important in determining the shape and amplitude of halo occupancies and whose underlying physics is clearest, rather than forcing ourselves to explain the behaviour of every HOD parameter in equal details and depth. Some of the parameters, especially those whose variations have little impact on the occupancy, have rather random and stochastic behaviours, of which we shall simply take note. This applies to both the ELG and the LRG fittings described below.

\subsubsection{Central galaxies}
\label{subsubsect:elg_cen_hod_evolution}

Let us consider the HOD parameters for central ELGs first. $\log M_c$ and $p_{\textrm{max}}$ are two of the central ELG HOD parameters which show the clearest model dependence and redshift evolution. According to Fig.~\ref{fig:parameter_vary_demo}, $\log M_c$ determines the peak position, at $M_{\rm p}\simeq10^{12} \,M_\odot$, of the central ELG occupancy, $\langle N_{\textrm{cen}}\rangle(M_h)$, while $p_{\rm max}$ controls the mean occupancy value at this peak. In GR, both parameters decrease over time, which is a result of smaller haloes starting star formation and forming ELGs later, and also of star formation in larger haloes becoming quenched. As more small haloes host ELGs, the peak position shifts leftward. Because smaller haloes are more abundant, the mean occupancy decreases for a fixed ELG number density. A decrease of $p_{\textrm{max}}$ -- and of $\sigma_M$ -- also helps to keep the ELG number density constant by narrowing the peak.

Because the HOD describes the connection between dark matter haloes and galaxies, it is affected by MG through the effects of the latter on both haloes and galaxies (the sSFR). An enhanced gravity helps make ELG-hosting haloes more massive, and increases the SFR, meaning that more haloes can start to house galaxies. Naively, this would increase both $M_c$ and $p_{\textrm{max}}$. However, Fig.~\ref{fig:ELG_fit_params} shows that $p_{\textrm{max}}$ is always lower in all MG models. This is because, as MG increases halo masses, it also increases the number of haloes near $M_{\rm p}$ and some of these haloes do not host ELGs, so the mean occupancy is decreased. The exception is when it is difficult to form new haloes of mass $\simeq M_{\text{p}}$, e.g., in emptier regions, while the enhanced gravity leads to stronger star formation, enabling more existing haloes to host ELGs (that meet our selection criteria). As an explicit check of this, we have split the ELG-hosting haloes into 4 bins of different environment density (cf.~Section \ref{sec:HOD environment effect}), and find that in the lowest-density environment $p_{\textrm{max}}$ for F6.0 is indeed increased. The total behaviour, however, is dominated by the high-density environments, leading to an overall smaller $p_{\textrm{max}}$ in F6.0 than in GR.

Based on the observations of Fig.~\ref{fig:ELG_fit_params}, we note an evolution sequence of the ELG central HOD and propose the following physical interpretation:
\begin{enumerate}
    \item \textit{initial unscreening}: ELG-hosting haloes become unscreened and experience faster growth in mass, leading to an increase of $M_c$ and decrease of $p_{\textrm{max}}$ compared with GR, as explained above.
    \item As time passes, some small haloes which initially do not host ELGs (that meet our selection criteria) have been unscreened for a while, and so experienced stronger star formation and started to have ELGs. As the star formation rate decays in the more massive haloes, ELGs tend to populate smaller haloes, decreasing $M_c$ to below the GR value. This crossover happens at redshift $1.0$, $0.7$, $0.4$ and $0.1$ for F4.5, F5.0, F5.5 and F6.0, respectively, clearly following a sequence. As smaller haloes are more abundant, and not all of them host a newly-added ELG, the mean ELG central occupancy at the peak further decreases, leading to a smaller $p_{\textrm{max}}$. However, we note that $p_{\textrm{max}}$ of F5.0 catches up with the GR value at $z=0$, while for F4.0 and F4.5 $p_{\textrm{max}}$ becomes larger than in GR between $z=0.1$--$0.2$. This is possibly because in these models the halo abundance at $M_h\simeq M_{\rm p}$ actually becomes smaller than for GR (see, e.g., the last two panels of Fig.~3 of Paper I) due to enhanced mergers.
\end{enumerate}

An interesting observation is that the different MG models tend to approach the same $M_c$ at late times. This suggests that our ELG selection criteria place a lower bound on the host halo mass, and haloes below this scale are unable to form ELGs (that meet our selection criteria) even with the boost by the fifth force.

There is a notable degeneracy between the $\sigma_M$ and $\gamma$ parameters, since they both control the width of the central occupation peak. See Fig.~\ref{fig:parameter_vary_demo} for a visual demonstration of this degeneracy. The result is that both parameters share a similar redshift and gravity dependence. The $\gamma$ fitting result appears to be noisier, with the fluctuations in redshift comparable to the model differences. 

The $Q$ parameter appears to be boosted by MG, possibly because the MG models induce more effective quenching in high mass haloes. 
However, a larger $Q$, like a smaller $p_{\textrm{max}}$ and a smaller $\sigma_M$, makes the peak of $\langle N_{\textrm{cen}}\rangle$ narrower on the right-hand side (cf.~Fig.~\ref{fig:parameter_vary_demo}), and so we expect there to be a degeneracy between these parameters, which makes the interpretation more complicated.

\subsubsection{Satellite galaxies}
\label{subsubsect:elg_sat_hod_evolution}

Next we discuss the fit parameters for the ELG satellites. All 3 satellite parameters appear to follow a similar evolution, which is most clearly seen in the $\log M_1$ fits. 

Based on Fig.~\ref{fig:HOD examples}, and supported by these results, we can identify the following evolution sequence of the satellite occupancy $\langle N_{\textrm{sat}}\rangle$ (since we are interested in the satellite population only, we will focus on $M_h\gtrsim10^{13}M_\odot$ where the HOD is dominated by satellites):
\textcolor{black}{
\begin{enumerate}
    \item screened regime -- where the deviation from GR is small because of chameleon screening. Example of this is F6.0 at $z=1.155$, where $M_1, M_{\textrm{cut}}$ and $\alpha$ all take values close to GR in Fig.~\ref{fig:ELG_fit_params}.
    \item initial unscreened regime -- small haloes become unscreened first in chameleon models. As unscreening starts, some haloes near $M_h=M_{\text{p}}$ go through enhanced star formation, and so their sSFR becomes higher than in some subhaloes in the more massive haloes (which are still screened) that host small satellites. Since our ELG selection criteria are based on ranked sSFR and we fixed the ELG number density $n_{\text{g}}$, the resulting catalogue in MG will have more centrals from smaller haloes and--correspondingly--fewer satellites from large haloes. This leads to a decrease of satellite occupancy in haloes more massive than $\simeq10^{13}M_\odot$. Examples of this include F5.5 and F5.0 at $z=1.155$, F6.0 and F5.5 at $z=0.652$ and F6.0 at $z=0.155$ in Fig.~\ref{fig:HOD examples}.\\ 
    \item catch-up regime -- the small central-hosting haloes enabled by the enhanced gravity in regime (ii) cannot sustain a continuous high sSFR: after some time their sSFR becomes lower and they are ``out-ranked'' once again by the satellites in massive haloes. These massive haloes, which saw a reduced satellite occupancy in regime (ii), now ``catch up" with the GR satellite occupancy. Examples of this regime are F4.5 at $z=1.155$, F5.0 at $z=0.652$ and F5.5 at $z=0.155$. We have checked this explanation explicitly by identifying pairs of matched haloes between the GR and F5.5 runs, and comparing their mean satellite occupancy at $z=1.155, 0.652, 0.155$ to find exactly the same sequence as described here.
    \item quenching regime -- similar to the halo growth, star formation rate also increases initially but decreases later, due to quenching or depletion. MG simply brings this process to earlier times. When haloes in an MG model have been completely unscreened for long enough, their star formation is suppressed, reducing the occupancy of ELG satellites (that meet our selection criteria). Examples of this are F4.0 at $z=1.155$, F4.0 and F4.5 at $z=0.652$ and F4.0, F4.5 and F5.0 at $z=0.155$.
\end{enumerate}
}

What we see in Fig.~\ref{fig:HOD examples} is that each model follows roughly this trend, though not all models go through all four regimes in the redshift range shown. For examples, F4.0 has entered regime (iv) before $z=1.155$, while F6.0 has not entered regime (iii) even at $z=0.155$. We have checked explicitly that this evolution sequence is not a numerical fluctuation in our fitting, by looking at more intermediate redshifts: Fig.~\ref{fig:ELG_HOD_evol_more_snapshots} demonstrates this using nine snapshots between $z=1.155$ and $z=0.05$.

This evolution sequence can also be seen in the satellite HOD parameters in Fig.~\ref{fig:ELG_fit_params}. Of the three parameters, $M_{\text{cut}}$ is the most poorly constrained, as its effect is mainly at low mass, where the satellites sub-dominate. Therefore here we focus on $M_1$ and $\alpha$, using Fig.~\ref{fig:ELG_HOD_evol_more_snapshots} to assist our visualisation:
\begin{enumerate}
    \item[F6.0:] regime (i) at $z=1.155$, starts to move towards regime (ii) at $z\simeq1$ and remains there until $z\simeq0$. In Fig.~\ref{fig:ELG_fit_params} its $M_1$ and $\alpha$ start to deviate from GR at $z\simeq1$.
    \item[F5.5:] starts in regime (ii) at $z=1.155$, moves towards regime (iii) between $z\simeq0.6$ and $z\simeq0.3$, and remains there until $z\simeq0$. In Fig.~\ref{fig:ELG_fit_params} its $M_1$ and $\alpha$ both attain a maximal deviation from GR at $z\simeq0.6$ and move to the GR values after that. 
    \item[F5.0:] starts in regime (ii) at $z=1.155$, moves towards regime (iii) between $z\simeq1.0$ and $z\simeq0.8$, remains there until $z\simeq0.4$ and by $z=0.155$ it has reached regime (iv). In Fig.~\ref{fig:ELG_fit_params} its $M_1$ and $\alpha$ values show maximal deviations from GR at $z=1.155$ and are closest to GR between $z\simeq0.8$ and $z\simeq0.5$.
    \item[F4.5:] starts in regime (iii) at $z=1.155$, moves towards regime (iv) between $z\simeq1.0$ and $z\simeq0.6$, and remains there ever since. In Fig.~\ref{fig:ELG_fit_params} its $M_1$ and $\alpha$ are close to GR at $z=1.155$ and catch up with the F4.0 values at $z\simeq0.6$.
    \item[F4.0:] already in regime (iv) at $z=1.155$ and stays therein ever since. In Fig.~\ref{fig:ELG_fit_params} its $M_1$ and $\alpha$ values are always very different from GR, with a lower $\alpha$ indicating a smaller satellite occupancy.
\end{enumerate}
We note that satellite occupancy data is quite noisy (cf.~Fig.~\ref{fig:HOD examples}) and does not seem to be always described by a perfect power-law function as Eq.~\eqref{eq:satellite_HOD_model}.

\subsection{Time evolution of LRG HOD}
\label{subsect:lrg_hod_evolution}

\begin{figure*}
    \centering
    \includegraphics[width=\linewidth]{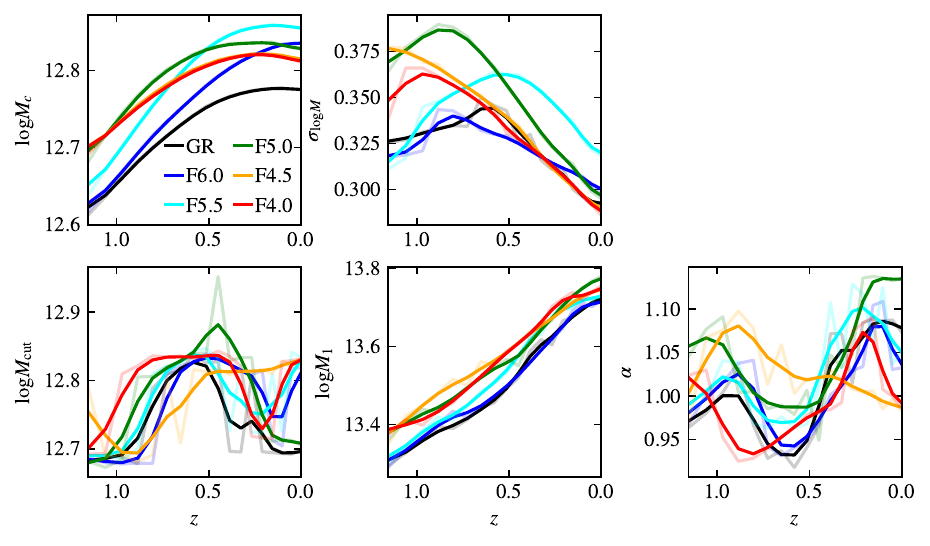}
    \caption{Best-fit LRG HOD parameters as a function of redshift. The HODs correspond to galaxy populations with a fixed number density of $0.001 \, h^{3}\mathrm{Mpc}^{-3}$. The solid curves show the best-fit parameters smoothed over redshift using Gaussian smoothing with $\sigma = 1.5$ plotted points. The unsmoothed curves are shown underneath as faded curves of the same colour.}
    \label{fig:LRG_fit_params}
\end{figure*}

The 5 LRG fit parameters are shown in Fig. \ref{fig:LRG_fit_params}. The parameters $M_c$ and $\sigma_{\log M}$ together control the LRG central galaxy occupancy $\langle N_{\textrm{cen}}\rangle$: $M_c$ represents the centre of the $\langle N_{\textrm{cen}}\rangle$ transition from $0$ to $1$, while $\sigma_{\log M}$ the width of the transition. From the GR results, we see that $M_c$ grows quickly and $\sigma_{\log M}$ increases at $z\gtrsim0.6$, while at $z\lesssim0.6$ $\sigma_{\log M}$ turns over and decreases and $M_c$ grows less quickly. This suggests that overall haloes grow faster at $z\gtrsim0.6$, causing the LRG-hosting haloes to very rapidly become more massive on average. During this period, more haloes below $M_c$ start to host LRGs (that meet our selection criteria), hence widening the transition region of $\langle N_{\textrm{cen}}\rangle$ from $0$ to $1$. At $z\lesssim0.6$, the growth of LRG-hosting haloes slows down, but this slowdown does not happen evenly across the board; instead, the more (less) massive ones slow down more (less), narrowing the transitional region of $\langle N_{\textrm{cen}}\rangle$. %This makes sense, since we have kept the number density of the LRGs constant over time, and it is likely that at low redshift there are few new haloes that start to host LRGs (that meet our selection criteria).

The MG effects on $M_c$ and $\sigma_{\log M}$ are non-trivial due to the complicated behaviour of the chameleon screening. Broadly speaking, we expect two effects. First, the fifth force enhances the growth of LRG-hosting haloes, increasing their mass. Second, it also enhances the formation and growth of galaxies, enabling new haloes to host LRGs (that meet our selection criteria). The first effect starts earlier, as haloes grow due to accretion from their (lower-density) surroundings, which become unscreened earlier than the (higher-density) inner regions where stars and galaxies form from cooled gas. For the number density of LRGs here, $M_c$ is in the region of $\simeq10^{12.7}M_\odot$, a relatively small mass at which haloes in models such as F4.0 and F4.5 have long become unscreened by $z\simeq1$ \citep{Mitchell:2018qrg}. This means that for these two models -- and to a lesser extent F5.0 as well -- both of the above two effects have had time to exert influence. In particular, the second effect means that in these models the fifth force can help more haloes with $M<M_c$ to host LRGs (that meet our selection criteria), hence widening $\sigma_{\log M}$ by $z\simeq 1$.

Two observations here are of potential interest. First, F4.0 and F4.5 have identical $\sigma_{\log M}$ to GR at $z\lesssim0.6$. They also have identical $\log M_c$ to each other within the entire redshift range shown in Fig.~\ref{fig:LRG_fit_params}, which is a constant shift compared with the GR result. These seem to imply that these models have the same population of haloes that host the selected LRGs, and that the masses of these haloes are enhanced by roughly the same factor in F4.0 and F4.5. This is possible, since in these two models the fifth force has become fully unscreened well above $z\simeq1$ (for the halo mass range of interest here), so that it has enhanced the growth of both haloes and the LRGs they host, keeping the ranking order of both unaltered. In other $f(R)$ variants, such as F6.0, haloes at $M\simeq M_c$ are still experiencing unscreening, so it is possible that some small haloes become unscreened and grow more massive than other, initially more massive, haloes. This changes the ranking orders of halo mass and stellar mass, complicating the halo-galaxy connection.

The second observation is that models such as F4.0, F4.5 and F5.0 all have earlier peaks of $\sigma_{\log M}(z)$, due to the fifth force bringing halo growth to earlier times, consistent with our explanation of the GR behaviour above.

On the other hand, the behaviour of $\sigma_{\log M}$ in F6.0 is typical of a case where unscreening happens later, and only the first MG effect mentioned earlier---halo mass enhancement---has had a chance to influence. Smaller haloes with $M<M_c$ have now experienced an enhanced growth in mass, while larger haloes with $M>M_c$ have not. The growth of stellar mass has not been affected by MG in either of these haloes. This causes a narrowing of the transition region of $\langle N_{\textrm{cen}}\rangle$, and hence a smaller $\sigma_{\log M}$. A similar behaviour can be seen for F5.5 at $z>1$, and we have checked that this is even clearer if looking at F5.5 haloes in high-density environments (which become unscreened later). We also note that at $z\lesssim0.2$ F6.0 has larger $\sigma_{\log M}$ than GR, catching up with the behaviour of the stronger MG models.

The satellite HOD parameters for LRGs are quite noisy, with the exception of $M_1$, which appears to be typically larger in stronger models. Again, here we note that F4.0 and F4.5 have nearly identical values at $z\lesssim0.6$, and F5.5 also approaches these two models at $z\simeq0.5$. This is consistent with our speculation above that for strong MG models such as F4.0 and F4.5, the LRG-hosting haloes are the same halo population as in GR but with universally enhanced masses. %\textcolor{red}{In Eq.~\eqref{eq:satellite_HOD_model}, we see that a constant shift of both $M_h$ and $M_1$ leaves the satellite occupancy number $\langle N_{\textrm{sat}}\rangle$ unchanged (given that $M_{\textrm{cut}}\ll M_1$ and that both $M_{\textrm{cut}}$ and $\alpha$ have little time/model variation).}
\section{Secondary HOD dependency: environment}
\label{sec:HOD environment effect}

Having considered the ensemble behaviour of the HOD parameters, we are now interested in secondary quantities which could affect the galaxy-halo connection, e.g., the assembly bias. In this section, we divide each galaxy catalogue into four ``sub-catalogues" of different environments by the following procedure:
\begin{enumerate}
    \item Separate galaxies into bins of host halo mass.
    \item Within each bin further divide the population into four quartiles of halo environment (see environment definition below), i.e. the haloes within each mass bin will be equally split into 4 bins.
    \item Combine galaxy contents of each environment quartile in all mass bins to form 4 galaxy sub-catalogues.
\end{enumerate}
The environment of a halo is characterised by the overdensity of the spherical annulus centred around the halo, with radii $2R_{200\textrm{m}}\leq r\leq 8\,\textrm{Mpc}$. We settled on this choice after experimenting with various definitions.
A fixed inner radius for the largest objects can include a significant amount of mass belonging to the halo itself in the environment definition. The (variable) inner radius should be somewhat larger than the virial radius and splashback radius for most objects. We chose not to scale the outer radius with $R_{200\textrm{m}}$, because a variable outer radius would mean that the environment measurements are not physically comparable between small and large haloes. Note that other definitions of environments are possible (and worth exploring in future), such as a smooth regular grid.

In what follows, we refer to the sub-catalogues as Q1-Q4 with Q1 being the lowest density quartile and Q4 the highest density one.

In studies of $\Lambda$CDM, it has been found that environment plays a significant role in galaxy formation \citep[e.g.][]{Hadzhiyska2020, Yuan_2021}. In $f(R)$ gravity, there can be an additional, inherent, environmental dependency thanks to the chameleon screening mechanism. For example, in $\Lambda$CDM, LRGs show a known preference for forming in higher density environments. ELGs are perhaps more interesting: in Paper I, we found large deviations of large-scale ELG clustering from GR in MG models, and suggested that an environment effect plays a role. The bulk of ELGs are central galaxies housed by haloes of mass $M_{\rm 200c} = 10^{12}$--$10^{13}\, M_\odot$. Because this is a relatively small halo mass range, the continuous screening of objects in dense regions to later times and the unscreening in empty regions can easily lead to a substantial environmental effect in MG models. One can end up in a situation where a majority of objects in overdense (underdense) environments are screened (unscreened).

\subsection{The environment dependence of ELG HOD}
\label{subsect:env_hod_elg}

With our galaxy catalogues divided into 4 environment quartiles as described above, we fit the HOD model within each quartile, starting with ELGs.

\begin{figure*}
    \centering
    \includegraphics[width=\linewidth]{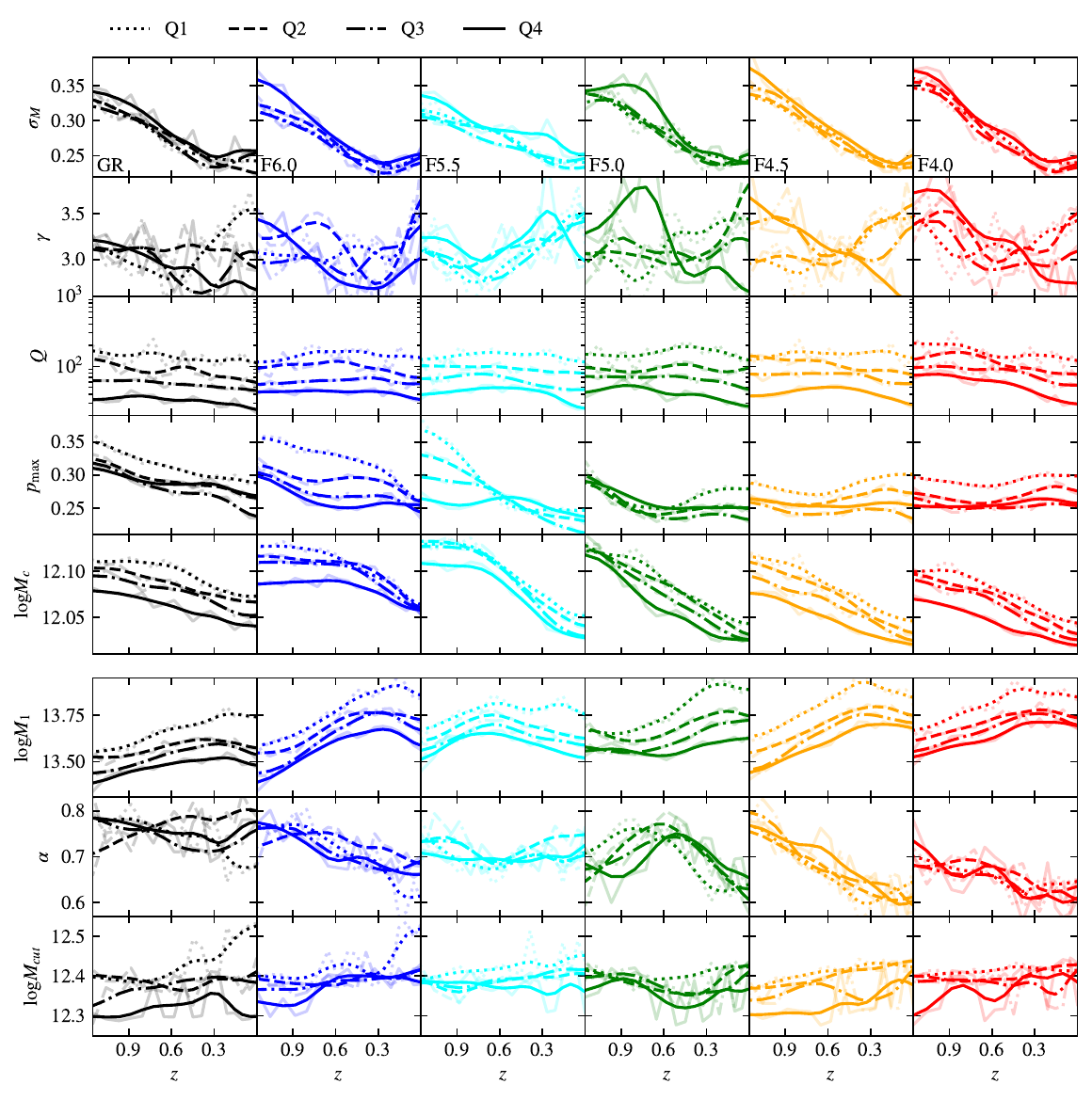}
    \caption{HOD parameters for ELG samples at $n_{\text{g}}=0.001\, h^{3}\mathrm{Mpc}^{-3}$, divided into 4 subsamples corresponding to galaxies hosted by haloes in a particular quartile of environment density. Each row corresponds to one  parameter, and each column corresponds to  parameter values of one gravity model, with line styles indicating different quartiles of environment. To improve visibility, solid curves show the fitted parameters smoothed over redshift using a Gaussian smoothing kernel with $\sigma = 1.5\, \rm points$, while faded curves underneath show the unsmoothed counterpart.
    Quartiles are labelled Q1-4 with Q1 (Q4) corresponding to the galaxies hosted by haloes in the least (most) dense environment.}
    \label{fig:ELG_HOD_env_fits}
\end{figure*}

In Figure \ref{fig:ELG_HOD_env_fits} we show the ELG HOD parameters as functions of redshift for each gravity model (per column). Most central parameters show fairly clean fit curves with the exception of $\gamma$, which is a less impactful parameter on the shape of the HOD. The parameters in different environment bins largely follow their overall behaviour, but there is a clear environmental dependence.

Again, we focus on $M_c$ and $p_{\text{max}}$, as these are the more important ELG central HOD parameters. Both parameters follow the same evolution sequence described in Section \ref{subsect:elg_hod_evolution}. We find that $M_c$ is larger in low-density environments for all models and at all redshifts, suggesting that ELGs form in more massive haloes in low-density environments. These haloes also usually have a higher mean central ELG occupancy at the peak, and hence higher $p_{\text{max}}$, which could be because they are less abundant. However, there are redshift ranges within which an MG model can have lower $p_{\textrm{max}}$ in low-density environments, the most notable example of which is F5.5 at $z\lesssim0.75$, but also F5.0 at $z\gtrsim0.5$. This is likely because the fifth force has strongly increased the number of haloes near $M_p$ at these redshifts. Actually, a closer inspection shows that, for F6.0, $p_{\text{max}}$ in the least dense environments is getting smaller than in the most dense environments right at $z\simeq0$, suggesting that this is yet another evolution sequence that is controlled by the unscreening of haloes. We shall see in Section \ref{subsect:ab_elg} that this has interesting consequences for the assembly bias of ELGs.

Now we look at the satellite ELG HOD parameters. The results for $\alpha$ and $M_{\text{cut}}$ are quite noisy, while $M_1$ has a clear environmental dependence, smaller in denser environments for all models and at all redshifts. A larger $M_1$ means smaller satellite occupancy, $\langle N_{\text{sat}}\rangle$, according to Eq.~\eqref{eq:satellite_HOD_model}, and so this suggests that haloes in less dense environments form fewer satellite ELGs on average than those in dense environments.

\begin{figure*}
    \centering
    \includegraphics[width=\linewidth]{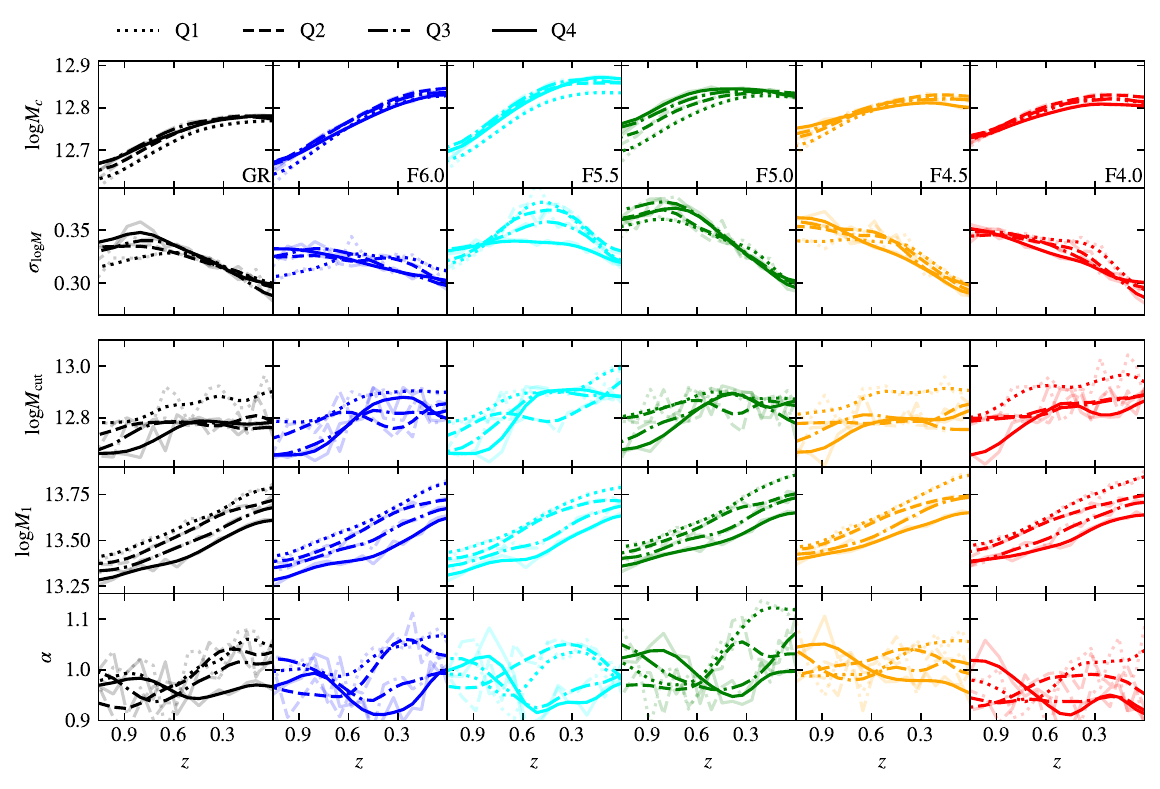}
    \caption{HOD parameters for LRG samples at $n_{\text{g}}=0.001\, h^{3}\mathrm{Mpc}^{-3}$, divided into 4 subsamples corresponding to galaxies hosted by haloes in a particular quartile of environment density. Each row corresponds to one  parameter, and each column corresponds to  parameter values of one gravity model, with line styles indicating different quartiles of environment. To improve visibility, solid curves show the fitted parameters smoothed over redshift using a Gaussian smoothing kernel with $\sigma = 1.5\, \rm points$, while faded curves underneath show the unsmoothed counterpart.
    Quartiles are labelled Q1-4 with Q1 (Q4) corresponding to the galaxies hosted by haloes in the least (most) dense environment.}
     \label{fig:LRG_HOD_env_fits}
\end{figure*}

\subsection{The environment dependence of LRG HOD}
\label{subsect:env_hod_lrg}

The LRG parameters in the environment quartile sub-catalogues are shown in Fig.~\ref{fig:LRG_HOD_env_fits}. There are clear differences in the environmental dependency for both $M_c$ and $\sigma_{\log M}$, but the detailed behaviour is complicated. In all environmental density bins, $M_c$ increases over time initially, but flattens and starts to decrease at lower redshift. As we explained in Section \ref{subsect:lrg_hod_evolution}, this represents the slowdown of the growth of haloes with $M_h\simeq M_c$. Haloes in the most dense environments tend to grow faster than those in the least dense ones, as can be seen from the GR results in Fig.~\ref{fig:LRG_HOD_env_fits}. The MG effect on the environmental dependence of $M_c$ is more complex: for models where unscreening is still happening, such as F6.0, haloes have been unscreened in the least dense environments but remain screened in the most dense ones, so that halo growth is boosted more, causing $M_c$ to be larger, in the former. For models such as F5.5 and F5.0, haloes in all environments have been unscreened for long enough so that those in dense environments again grow more and have larger $M_c$. %\textcolor{orange}{For even stronger models, the $M_c$ peak happens earlier in less dense environments, such as F4.5. In F4.0, $M_c$ is consistently lowest in the least dense environments, which can be because more small haloes have been able to form LRGs, widening the transition of $\langle N_{\textrm{cen}}\rangle$ between 0 and 1.}

Similarly, for $\sigma_{\log M}$, initially it is larger in dense environments, suggesting that in such environments more haloes with mass somehow below $M_c$ can host LRGs. In other words, haloes form galaxies more easily in denser environments, where the requirement on halo mass can be relaxed. The $\sigma_{\log M}$ value in the least dense environments tends to catch up with, and sometimes exceed, the value in denser environments, e.g., F6.0 at $z\simeq0.5$ and F5.5 at $z\simeq0.8$. In all models the $\sigma_{\log M}$ values in different environments tend to converge at low redshifts, consistent with the observation in Fig.~\ref{fig:LRG_fit_params} that $\sigma_{\log M}$ for all models converge to the GR value at low $z$.

Of the satellite parameters, we again see that $M_1$ has the cleanest fit, while $M_{\rm cut}$ and $\alpha$ curves are noisier. As in the ELG case of Fig.~\ref{fig:ELG_fit_params}, $M_1$ is smaller in denser environments, at all redshifts and for all models, signalling that haloes in dense environments host more satellites on average, which is reasonable since we expect that there are more subhaloes in these environments that host LRGs and have fallen into them.
\section{HOD dependency: concentration}
\label{sec:HOD concentration effect}

We follow a similar procedure to divide up our galaxy population as described in Section \ref{sec:HOD environment effect}, but instead dividing the halo population into quartiles of different halo concentrations, defined here as $c_{200c} = R_{200c}/R_s$ where $R_s$ is the scale radius of the NFW profile, within each mass bin. $c_{200c}$ values were obtained via fitting the NFW profile to the halo dark matter density profiles of the dark matter haloes. 

The formation of galaxies is expected to have a connection to host halo concentration, as more concentrated haloes have a steeper density profile and thus should be more efficient at cooling gas. This makes concentration an interesting choice of second parameter for the HOD. \cite{Mitchell2019} performed an in depth investigation of the impact of the MG effect on halo concentration, and proposed a universal model of this effect for $f(R)$ gravity. Haloes below the ``screening mass", $10^{p_2}$, with $p_2$ given in \cite{Mitchell2019}, have up to $40\%$ greater median concentration than GR objects in the same mass bin. This motivates our choice to investigate this dependency in this section.

\subsection{The concentration dependence of ELG HOD}
\label{subsect:concentration_hod_elg}

\begin{figure*}
    \centering
    \includegraphics[width=\linewidth]{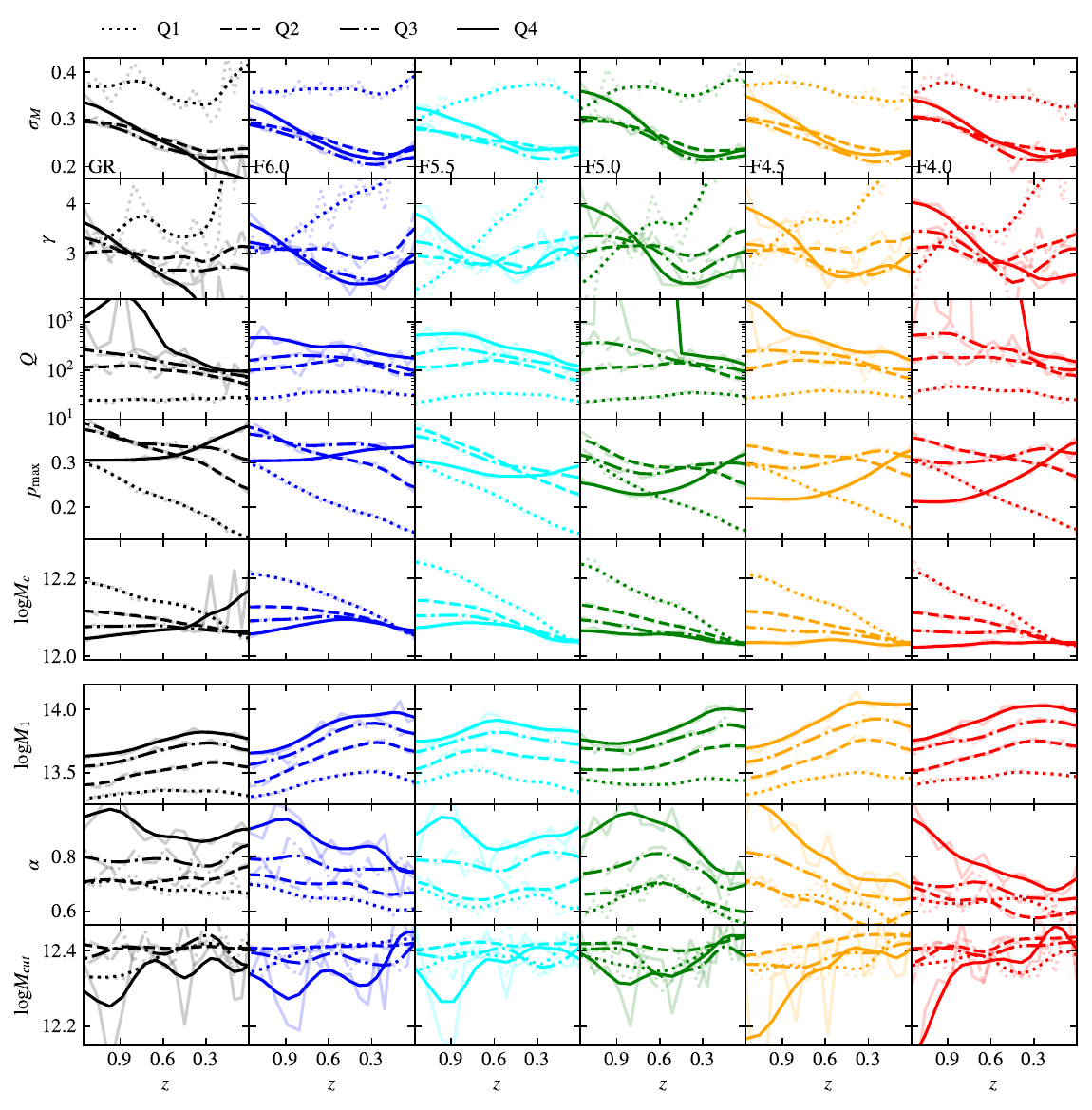}
    \caption{HOD parameters for ELG samples at $n_{\text{g}}=0.001\, h^{3}\mathrm{Mpc}^{-3}$, divided into 4 subsamples corresponding to galaxies hosted by haloes in a particular quartile of concentration. Each row corresponds to one parameter, and each column corresponds to fitted parameter values of one gravity model, with line styles indicating different quartiles of concentration. To improve visibility, solid curves show the  parameters smoothed over redshift using a Gaussian smoothing kernel with $\sigma = 1.5\, \rm points$, while faded curves underneath show the unsmoothed counterpart. Quartiles are labelled Q1-4 with Q1 (Q4) corresponding to the galaxies occupying haloes with the lowest (highest) concentration.}
    \label{fig:ELG_HOD_concentration_fits}
\end{figure*}

In Fig.~\ref{fig:ELG_HOD_concentration_fits} we show the ELG HOD fit parameters in quartiles of concentration and their redshift evolution. We start with an initial observation that the typical \textcolor{black}{variation across the} quartiles is greater than what is shown in Fig.~\ref{fig:ELG_HOD_env_fits}, indicating that the HOD has a stronger concentration dependence than they do environmental dependence. We will begin by analysing the central ELG parameters.

For all models, $M_c$ monotonically decreases as the concentration $c$ increases. Less concentrated haloes are less efficient in cooling gas and triggering star formation, which explains why they have to be more massive (thus the larger $M_c$) to start hosting ELGs. $M_c$ in the lowest-$c$ bin decreases rapidly with time, suggesting that less concentrated haloes can still form ELGs, but need more time. $M_c$ decreases less rapidly for higher-$c$ bins---for the highest-$c$ bin it remains nearly constant---over time. At low redshift the difference in $M_c$ between the different concentration bins disappears.

The behaviour of $p_{\text{max}}$ is more complicated. In most cases it is lowest for the lowest-$c$ bin, and decreases monotonically over time. This indicates that least concentrated haloes are less likely to host a central ELG, and it is even more so at later times. Haloes in the other concentration bins are much more likely to host ELGs on average, but the dependence of $p_{\text{max}}$ on $c$ is not monotonic: in all models the most concentrated haloes have smaller $p_{\text{max}}$ than haloes in the two intermediate-$c$ bins until low $z$. In the stronger models, F5.0, F4.5 and F4.0, $p_{\text{max}}$ in the highest-$c$ bin is even smaller than in the lowest-$c$ bin until $z\simeq0.6$, implying that the most concentrated haloes are less likely to host a central ELG than the least concentrated ones. A clear physical explanation is hard to come up with as in practice all the HOD parameters vary simultaneously and their combined effect is not straightforward to predict, but a potential reason is that the more concentrated haloes have black hole growth earlier, triggering the AGN which quenches the galaxies earlier. We note that the behaviour of $p_{\text{max}}$ seems to ``mirror'' that of $\sigma_M$, i.e., the lowest-$c$ bin has the largest $\sigma_M$ and the highest-$c$ bin also tends to have larger $\sigma_M$. It is therefore possible that, for the highest-$c$ bins, the effect of a smaller $p_{\text{max}}$ (lower peak height of $\langle N_{\text{cen}}\rangle$) is compensated by that of a larger $\sigma_M$ (a larger peak width). A more detailed investigation into this is beyond the scope of this paper. We note in passing that the lowest-$c$ bin has consistently the smallest $Q$ which, like a large $\sigma_M$, also implies a larger peak width of $\langle N_{\text{cen}}\rangle$.

We next briefly discuss satellite parameters. We notice, again, that $M_{\text{cut}}$ is too noisy to discern trends from. $M_1$ and $\alpha$ are both larger in the highest-$c$ bins, implying simultaneously a smaller normalisation and a steeper slope of the satellite occupancy, $\langle N_{\text{sat}}\rangle(M_h)$, for the most concentrated haloes. The trend is clear and the effect is strong. We suspect that tidal stripping plays a role here: smaller haloes are more concentrated on average and have fewer satellites; the more concentrated ones of them experience stronger tidal stripping, an effect which is made even more noticeable by the fact that there are very few (e.g., $<1$ per halo) satellites to start with. Larger haloes, on the other hand, contain more satellites (up to $\simeq20$ in our catalogues) and are less concentrated, such that even the ones with the highest concentrations still have a milder tidal stripping, and so the overall effect on the satellite number is smaller. Together these make $\langle N_{\text{sat}}\rangle(M_h)$ steeper. Again, a more detailed exploration of the physics behind this observation will be left for future work.

\subsection{The concentration dependence of LRG HOD}
\label{subsect:concentration_hod_lrg}

\begin{figure*}
    \centering
    \includegraphics[width=\linewidth]{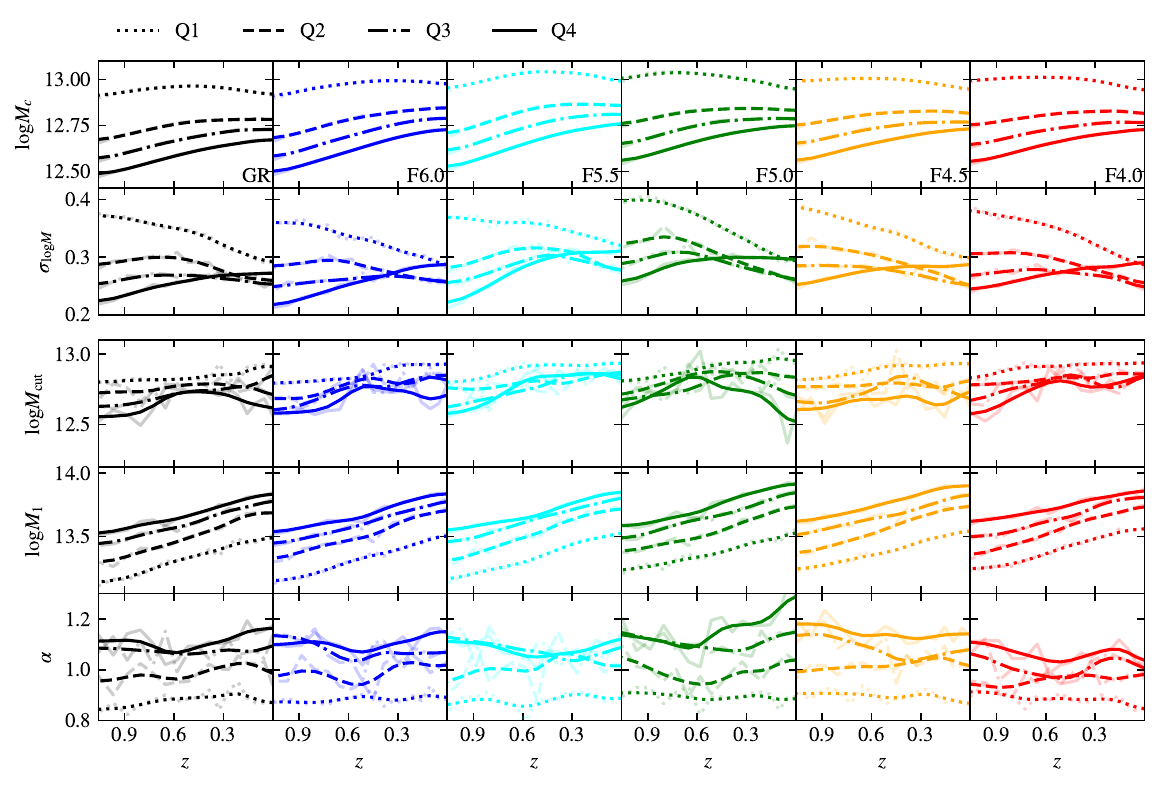}
    \caption{HOD parameters for LRG samples at $n_{\text{g}}=0.001\, h^{3}\mathrm{Mpc}^{-3}$, divided into 4 subsamples corresponding to galaxies hosted by haloes in a particular quartile of concentration. Each row corresponds to one parameter, and each column corresponds to fitted parameter values of one gravity model, with line styles indicating different quartiles of concentration. To improve visibility, solid curves show the  parameters smoothed over redshift using a Gaussian smoothing kernel with $\sigma = 1.5\, \rm points$, while faded curves underneath show the unsmoothed counterpart. Quartiles are labelled Q1-4 with Q1 (Q4) corresponding to the galaxies occupying haloes with the lowest (highest) concentration.}
    \label{fig:LRG_HOD_concentration_fits}
\end{figure*}

In Fig.\ref{fig:LRG_HOD_concentration_fits} we present the LRG HOD parameters for haloes in different quartiles of concentration. Similar to with the ELGs, we see a much greater dependence on concentration than on environment for the HOD. The behaviour of the satellite parameters are similar to that of ELG satellites -- larger $M_1$ and $\alpha$ for higher-concentration bins -- and the explanation given in Section \ref{subsect:concentration_hod_elg} can be applied here.

The parameters for the central LRG HOD, $M_c$ and $\sigma_{\log M}$, also show a clear and clean dependence on the halo concentration $c$. It is harder for less concentrated haloes to form central galaxies, and this is reflected by the larger $M_c$, i.e., less concentrated haloes tend to be more massive \textit{on average} when they form central LRGs. This is compensated by a wider transition region of $\langle N_{\text{cen}}\rangle(M_h)$, i.e., a larger $\sigma_{\log M}$, which indicates that there is a larger variety of the LRG-hosting halo mass for the lower-$c$ bins.

\section{Assembly bias}
\label{sect:ab}

Pipelines for full-shape galaxy clustering analyses  in cosmological surveys usually model the galaxy-halo connection with simple HOD models, ignoring any other effects. However, we see above that the growth history of both haloes and galaxies have a strong dependence on their environment, which suggests that assembly bias may have an impact on the galaxy distribution and clustering at late times.

Here, we assess quantitatively the impact of assembly bias on the two-point galaxy correlation function in MG models to see if this impact depends on the gravity theory. Since the correlation function is a function of galaxy separation, to make this task more tractable, we compress it into one number -- the large-scale correlation function (LSCF), $\xi_{\rm LS}$, introduced in Paper I. $\xi_{\rm LS}$ is defined as the volume-averaged $\xi$ in the range $r\in[5,20]~\textrm{Mpc}$. 

To this end, we have generated HOD catalogues of ELGs and LRGs using not only the basic HOD model, but also two variations which, respectively, incorporate the dependence of the HOD parameters on the environment densities and concentrations of host haloes. We directly use the measured HODs from our galaxy samples to assign galaxies to haloes, which avoids potential systematic and stochastic effects related to the fitting of HOD to the AB signal. Following the discussion of the assumptions of the basic HOD model in Sect.~\ref{subsec:assumptions_basic_HOD}, halo satellite occupation numbers are drawn from the distribution of the true samples within each bin of halo mass (and any secondary HOD parameter). Satellite positions are drawn from the radial distribution of satellites measured from the simulated galaxy catalogues (this distribution is measured as a function of $r/R_{\rm 200m}$) in 5 wide bins of halo mass. When considering the HOD including secondary parameters, the HOD is divided, within each halo mass bin, into a further 8 bins of that parameter, which are constructed within each mass bin by equally dividing the halo population. We calculate $\xi_{\textrm{LS}}$ for all three types of HOD, labelled as $\xi_{\rm gen}$ in Fig.~\ref{fig:generated correlation function}, and compare with the measured $\xi_{\textrm{LS}}$ from the hydrodynamical simulations directly, for all gravity models and for both ELGs and LRGs.

\subsection{The assembly bias of ELGs}
\label{subsect:ab_elg}

\begin{figure*}
    \centering
    \includegraphics[width= \linewidth]{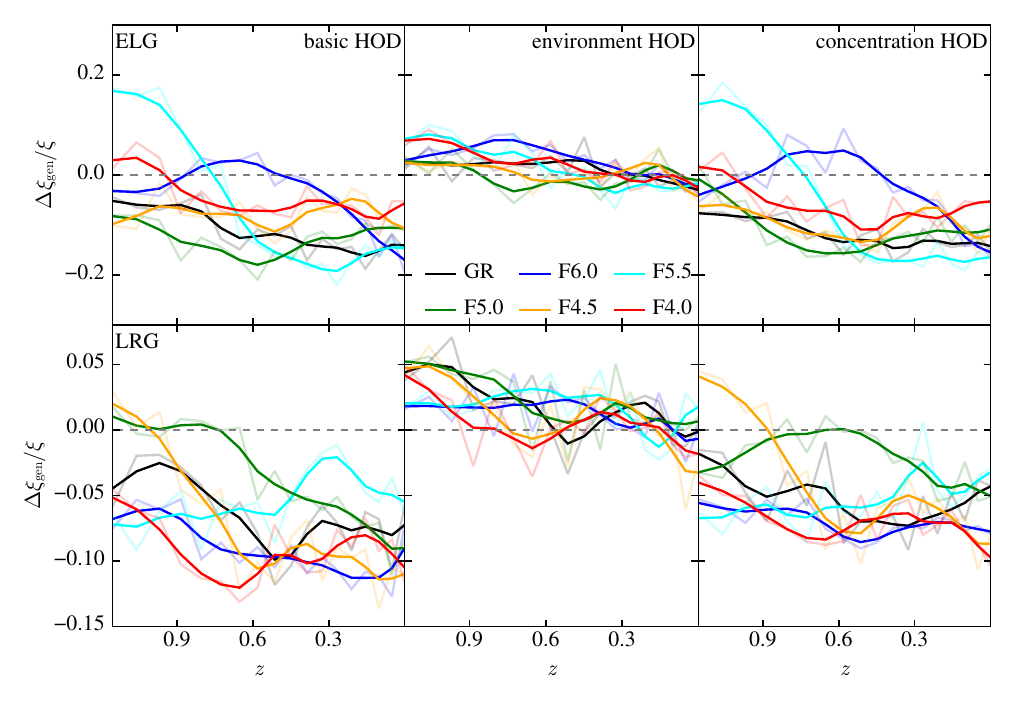}
    \caption{The averaged large-scale correlation function (defined in the main text) of ELG (top row) and LRG (bottom row) galaxy samples generated, from left to right panels, by:
    (i) using the measured basic HOD to populate haloes,
    (ii) using the measured HOD in environment percentiles to populate haloes, and
    (iii) using measured HOD in concentration percentiles to populate haloes. We show the relative difference between the LSCFs in the generated and simulated galaxy catalogues, using the latter as a measure of the ground truth. The same colour scheme denoting different models is used as before, and as shown by the legends. To increase visibility, we smooth the curves over redshift with a Gaussian smoothing kernel ($\sigma = 1.5 \, \rm$ data points), while the unsmoothed curves are shown as faded lines underneath.}
    \label{fig:generated correlation function}
\end{figure*}

In Fig.~\ref{fig:generated correlation function} we compare the LSCF of the generated samples. First we look at the ELG results, shown in the top row. The left panel shows the result of the basic HOD, where assembly bias is ignored entirely when generating HOD galaxy samples. As a result, deviations from the actual measurement of corresponding hydro simulations can be considered as an indicator of the impact of assembly bias.

We performed a rudimentary comparison of our ELG assembly bias signals to those found in TNG300. We found for both TNG300 and our simulations that increasing the stellar mass cut in the  ELG samples also strengthened the amplitude of the assembly bias. We tested this by comparing the fractional differences between the true samples and samples with galaxies shuffled within halo-mass bins. As a result,  we do expect to see a slightly higher assembly bias in our ELG samples than elsewhere in the literature.

There is clearly a sizeable AB effect across all models. In most cases, $\xi_{\textrm{LS}}$ of the generated samples is less than the prediction of hydrodynamical simulations, indicating that for a given halo mass bin galaxies typically prefer to occupy a subset of haloes that are more clustered than the whole population. However, there is a lack of order in the AB effect amongst the models and time snapshots. For example, the strongest models, F4.0 and F4.5, broadly show a weaker AB effect than in GR across the entire range of redshifts, while the strongest effect happens in a relatively weak model, F5.5. This overall behaviour implies that, for any given selection criteria, different models have different galaxy populations, which has a stronger effect in the clustering strength than modified gravity itself.

While such a complicated behaviour makes it difficult to fully identify the underlying physics, we can at least get some clues from the environmental dependence of the HOD parameters shown in Fig.~\ref{fig:ELG_HOD_env_fits}. Taking F5.5 as example, at higher $z$ there is a strong variation of $p_{\textrm{max}}$ in different environments, which is smaller in higher-density environments. As discussed above, this parameter has a strong impact on the HOD population, with a smaller $p_{\textrm{max}}$ implying a lower mean occupancy within the mass range $12.0\lesssim\log_{10}\left(M_\textrm{200c}/M_{\odot}\right)\lesssim12.5$. In $f(R)$ gravity, the chameleon screening mechanism means that objects in lower-density environments become unscreened first. In Paper I we suggested that the increased rate of accretion of gas onto unscreened objects leads to an earlier star formation, and thus an earlier seeding of ELG galaxies. We saw in Figure \ref{fig:ELG_HOD_env_fits} that there is indeed substantially enhanced central occupations in lower density environments for models such as F5.5 just after the mass range of central galaxies has become unscreened. These ELG-hosting haloes in low-density environments have grown from lower initial density peaks, and as a result are less clustered. On the other hand, the ELG-hosting haloes in dense environments are screened and their clustering strength remains similar to that in GR. As a result, the overall clustering of ELGs decreases (increases) when this environment dependency is (is not) taken into account. This means that $\xi_{\text{gen}}$, which is generated using the basic HOD that neglects any environmental dependence, is larger than the $\xi$ measured from simulations, which is what we see in the F5.5 curve at $z\gtrsim0.75$. At $z\sim0.75$, the ranking order of $p_{\textrm{max}}$ with environment density is reverted (see Fig.~\ref{fig:ELG_HOD_env_fits}), and correspondingly the F5.5 curve crosses $0$ in the upper left panel of Fig.~\ref{fig:generated correlation function}.

The above explanation based on the environmental dependence of $p_{\text{max}}$ seems to also work for F6.0, which has a very strong environmental difference at $z\gtrsim0.3$, and correspondingly in Fig.~\ref{fig:generated correlation function} its curve is closer to $0$ and above $0$ between $z\simeq0.9$ and $z\simeq0.5$, like F5.5. $p_{\text{max}}$ in both F4.5 and F4.0 has a similar level of environment dependence to GR, and the AB effects are also comparable in these models. F5.5 is an outlier, the only model where $p_{\text{max}}$ is smaller in low-density environments than in high-density ones, and this gives it the strongest AB effect in Fig.~\ref{fig:generated correlation function}.
%For the other $f(R)$ models (F5.0, F4.5, F4.0), $p_{\textrm{max}}$ is smaller than in GR for all environment bins, which is the result of enhanced halo abundances, but the behaviour of the curves in the upper left panel of Fig.~\ref{fig:generated correlation function} is harder to explain, because the difference of $p_{\textrm{max}}$ between the different environment bins is much smaller and has little variation, so that the MG effects on other HOD parameters may no longer be subdominant. 
Note that, in F4.5 and F4.0, the MG effect has `saturated' (as discussed above), so that $p_{\textrm{max}}$ in Fig.~\ref{fig:ELG_HOD_env_fits} and $\Delta\xi_{\textrm{gen}}/\xi$ in Fig.~\ref{fig:generated correlation function} behave similarly in these two models.

With this discussion, we look at our samples generated with the HOD measured in environment percentiles. For all these samples, the LSCF is now very close to the actual LSCF. This confirms that for GR and our considered MG models a majority of the assembly bias can be explained by the environment dependency of the HOD. We note that in other works \citep[e.g.,][]{Hadzhiyska2021}  the tidal environment has shown to be better at accounting for ELG assembly bias, particularly in the one-halo regime. Another possibility is to define the environment using the smoothed matter density on a regular grid. We have not considered these definitions here, since they are beyond the scope of this paper. Furthermore, $f(R)$-specific environmental effects are understood to be more closely related to the overdensity, rather than tidal properties, of the surrounding region.

There is some residual assembly bias, which is below $2$-$3\%$ at $z\lesssim0.5$ and up to about $6\%$ at higher redshift, in the environment HOD samples. Apart from statistical noise or stochasticity due to the choices of random seeds used to generate the HOD catalogues\footnote{We have tested the latter possibility by repeating this exercise using different HOD random seeds and found their effect too small to be the leading cause here. This indicates that the improvement of AB-effect modelling using environment HOD in Fig.~\ref{fig:generated correlation function} is real and not due to HOD seed variation.}, this can be due to a number of factors, including: (1) the environment definition does not perfectly account for the assembly bias, (2) for MG models the MG-specific environmental effect requires a slightly different prescription to GR, (3) there may be small dependency on additional physical quantities such as halo concentration or formation time.

To check the last possibility, next we draw our attention to the samples generated with the HOD measured in concentration quartiles (the upper right panel of Fig.~\ref{fig:generated correlation function}). The LSCFs are now visibly similar to those of the basic-HOD generated samples. 
Despite the HOD showing a substantial concentration dependency in Fig.~\ref{fig:ELG_HOD_concentration_fits}, the actual relevance of that for galaxy clustering is small, even in MG models. This suggests that the correlation between the AB and halo concentration is not a signature of causal relationship. This finding is consistent with results in the literature that concentration is not a strong candidate for an HOD parameter for the purposes of reproducing ELG clustering \citep[e.g.][]{Hadzhiyska2021, Yuan2022}.

\subsection{The assembly bias of LRGs}
\label{subsect:ab_lrg}

We compare generated and true LSCFs for LRGs in the bottom row of Fig.~\ref{fig:generated correlation function}. The strength of the AB is overall weaker than in the ELG case (note the different range of the vertical axis), and the variation between models is smaller. We note that the level of LRG assembly bias we find is smaller than in some other works. Using a smaller radial range (e.g., $2$--$10 \,\mathrm{Mpc}$) gives us level of assembly bias of around $10\%$, in concordance with the literature. After this radius range the difference between the true and generated sample CF decreases, and so our choice of range when defining the LSCF leads to this smaller assembly bias estimate. 

There is an increasing assembly bias over time for all models, and in most MG models the AB effect is stronger than in GR, with the only exception being F5.0 (and F4.5 at $z\gtrsim0.9$) which is likely because the effects of assembly bias and environmentally-dependent modified gravity cancel each other. As in the ELG cases, the MG effects on galaxy assembly bias are typically non-monotonic.

For environment-HOD generated samples (the bottom middle panel of Fig.~\ref{fig:generated correlation function}), we can see that for all models the LSCF is much closer to the true value, suggesting that again a majority of the LRG assembly bias is associated with the environment dependency of the HOD. The size of the AB effect is reduced to within $5\%$ at $z\gtrsim0.9$ and $2$-$3\%$ at $z\lesssim0.3$, in line with the ELG result. For GR this is also in line with the findings in other works \citep[e.g.][]{Hadzhiyska2020} and it is unsurprising that the same applies to MG models.

Lastly we consider concentration-HOD generated samples (the lower right panel of Fig.~\ref{fig:generated correlation function}). Concentration is clearly a less impactful secondary HOD variable than environment, as was the case with ELGs. 

The fact that halo concentration strongly affects galaxy formation (cf.~Figs.~\ref{fig:ELG_HOD_concentration_fits} and \ref{fig:LRG_HOD_concentration_fits}) but seems largely irrelevant for the assembly bias on the scales considered here (cf.~Fig.~\ref{fig:generated correlation function}) suggests that all concentration bins have roughly the same distribution of halo environment, which is the secondary halo property that is more relevant for assembly bias.

\subsection{Observational implications}

Overall, there is assembly bias of both LRGs and ELGs in the whole redshift range considered, where modern cosmological surveys will collect data. For some models and redshifts this assembly bias is larger in magnitude than in GR or even has an opposite sign. This 
can be of significance in designing tests of gravity using cosmological survey data. It is likely that it is insufficient to use the common basic HOD models without introducing biases in inferred cosmological parameters. This is even -- and at times particularly -- true for weaker MG models such as F5.5 and F6.0, which have very different ELG AB properties to GR at redshifts relevant in modern cosmological surveys such as DESI and \textit{Euclid}. 

We have shown here that, for all models and redshifts, the AB effect can be largely eliminated by considering a slight extension of the basic HOD model by introducing an environment dependency. This is convenient since the same environment definition is strongly relevant to the MG effects, in particular chameleon screening which is featured in $f(R)$ gravity.

Introducing such environment dependency does not fully get rid of the AB effect in the CF modelling, leaving a residual of {$\lesssim3\%$ ($\lesssim6\%$)} at low (high) redshift. It remains interesting to see how this level of assembly bias, if unaccounted for, might bias cosmological parameter inference, though this is beyond the scope of this study. It is possible that it may be necessary to use even more sophisticated HOD models. Again this will be left for future investigation.

\section{Discussion and Conclusions}
\label{sec:conc_and_disc}

Modern large-scale galaxy surveys, such as DESI, can measure the redshifts and positions of tens of millions of galaxies, allowing us to use galaxy clustering to constrain cosmological models and parameters. One of the most important theoretical uncertainties, however, is the galaxy-halo connection: how galaxies populate dark matter haloes and trace the underlying matter density field. This is because, while galaxies are what we observe, it is the matter density field, which is dominated by dark invisible matter, that directly responds to the cosmological model. Most galaxy clustering analyses nowadays assume the highly-flexible basic HOD model, which assumes that this connection depends only on halo mass. However, it has been widely known that haloes have assembly bias -- their formation or assembly history has an additional effect in their distribution. Comparisons of galaxy clustering based on the basic HOD model with the prediction of hydrodynamical simulations show that neglecting this assembly bias leads to weaker clustering strength in the prediction for $\Lambda$CDM, compared with the ground truth, for different types of galaxies. The AB effect is usually incorporated in the modelling by introducing secondary halo properties, such as the environmental density, in extensions to the basic HOD model.

The issue of assembly bias is more complicated in theories of modified gravity, where both halo evolution and galaxy formation -- e.g., via the effect on stellar mass or star formation rate -- can be affected. Furthermore, many MG models feature a screening mechanism, such as chameleon screening in the $f(R)$ models considered here, which can introduce further dependencies of galaxy formation on halo mass and environmental density. As we have seen throughout this paper, these effects work together in a complicated manner\footnote{It is worth pointing out that all the analyses in this paper are based on ELG and LRG catalogues at a fixed number density $n_{\text{g}}=0.001\,h^3\text{Mpc}^{-3}$. For galaxy catalogues with different number densities or galaxy types, the results may well be different. For example, a galaxy catalogue with a higher number density will involve smaller haloes, which become unscreened earlier, so the effect of MG can be different.}, making the MG effect on galaxy-halo connection extremely difficult to predict without realistic hydrodynamical simulations.

In this paper we studied the effects of modified gravity on the formation of ELG and LRG galaxies, which are relevant for large-scale modern cosmological surveys such as DESI and \textit{Euclid}, based on the first suite of cosmological galaxy formation simulations in MG theories \citep{Mitchell_2022}. 

By fitting the basic HOD model to the galaxy catalogues, and looking at the time evolution of the fit parameters (see Fig.~\ref{fig:ELG_fit_params} and Fig.~\ref{fig:LRG_fit_params}), we identify two interesting physical results. The first is that the galaxy-halo connection follows a clear evolutionary sequence. Different MG models go through the same sequence at different times, and at any given time different models are at different stages. This sequence can be explained by the unscreening of the fifth force, and its subsequent effects on halo growth and star formation (as well as their interplay). We believe this is the first time MG physics is analysed in depth in the context of galaxy formation.

The second is a saturation effect, first observed in Paper I in the context of galaxy formation/clustering in MG models. As the MG effect accumulates for longer, the impact on galaxy formation seems to be limited, and we observe that models F5.0 and F4.5 `catch up' with the strongest MG model, F4.0, for several quantities. We argue that this effect is physical, because for any given galaxy selection criteria, the Universe is unable to keep forming more of them indefinitely: the selected galaxies in different models largely form from the same sites in the simulation's initial conditions. MG, with its enhancement of the strength of gravity, mainly makes galaxies form earlier. One caution is that the above statement depends on the simulation resolution: low-resolution simulations are fundamentally unable to form small galaxies. However, we have carefully chosen our galaxy cuts so that our results are not affected by limited resolution.

In the later part of this paper we have focused on the dependency of the HOD on secondary halo properties and assembly bias. Using hydrodynamical simulations, we have explored two secondary halo properties: their environmental density (Figs.~\ref{fig:ELG_HOD_env_fits} and \ref{fig:LRG_HOD_env_fits}) and concentration (Figs. ~\ref{fig:ELG_HOD_concentration_fits} and \ref{fig:LRG_HOD_concentration_fits}). 
We find substantial HOD parameter dependency on both environment and concentration in all models including standard $\Lambda$CDM, with the dependency being stronger than in $\Lambda$CDM in certain MG models. In line with other works in the literature, we find that including the concentration dependency in an extended HOD model does not noticeably get rid of the AB effect in the galaxy clustering prediction, but including the environment dependency does. Our results in Fig.~\ref{fig:generated correlation function} show that halo environment accounts for most of the assembly bias for both ELGs and LRGs. 
Furthermore, in the context of modified gravity, we observe novel assembly bias effects, particularly for ELGs, where the magnitude and direction of the assembly bias effect can be significantly different to $\Lambda$CDM. This is especially true for weaker MG models such as F5.5 (Fig.~\ref{fig:generated correlation function}).

We conclude that using basic HOD models for cosmological tests of modified gravity, especially those involving ELGs, may not be appropriate and may bias estimates of cosmological parameters or lead to misidentification of the gravity model. This is particularly important for modern surveys such as \textit{Euclid} and DESI, which probe clustering to smaller scales, where assembly bias may be stronger. 

The simple extension to the basic HOD by including the environmental effect through four environment density bins enables us to reduce the AB effect on ELG and LRG clustering to as small as {$2\%$}. A further elimination of the AB effect may involve more sophisticated treatment, possibly by including other secondary halo properties in the HOD model, or using fit parameters measured in a larger number of percentiles of environment densities. Our simulation box volume is not large enough to realistically allow for the latter to be done. This underscores the need for further work using larger and higher-resolution hydrodynamical simulations of MG, which could be used to produce more realistic ELG populations and other galaxy types such as bright galaxies. We caution that for galaxy formation simulations, higher resolution does not necessarily mean more realistic and we believe that simulation resolution may affect our results quantitatively but will not change the main qualitative conclusions. Nevertheless, higher simulation resolution will indeed allow us to do convergence tests of various quantities and produce galaxy catalogues with higher number densities.

%Results could be more relevant for different types of statistics, e.g. ones more related to smaller-scale clustering            

\section*{Acknowledgements}

This work used the DiRAC@Durham facility managed by the Institute for Computational Cosmology on behalf of the STFC DiRAC HPC Facility (https://www.dirac.ac.uk). The equipment was funded by BEIS capital funding via STFC capital grants ST/K00042X/1, ST/P002293/1, ST/R002371/1 and ST/S002502/1, Durham University and STFC operations grant ST/R000832/1. DiRAC is part of the National e-Infrastructure. 

MC is supported by a UK Science and Technology Facilities Council (STFC) PhD studentship. SB is supported by the UK Research and Innovation (UKRI) Future Leaders Fellowship (Grant No.~MR/V023381/1 and UKRI2044). BL is supported by the European Research Council (ERC) Advanced Grant `UNCA', under the UKRI's Frontiers Research Guarantee No.~EP/Z533877/1, and the UK STFC Consolidated Grant No.~ST/X001075/1, and acknowledges the hosts by the National Astronomical Observatory of China in 2025/2026, during which part of this work was carried out.

\section*{Data Availability}

Post-processed simulation data described in this paper can be requested by contacting the authors. Raw data of the simulations used in this work may also be requested from the authors.

%%%%%%%%%%%%%%%%%%%%%%%%%%%%%%%%%%%%%%%%%%%%%%%%%%



\bibliographystyle{mnras}
\bibliography{References} % if your bibtex file is called example.bib

%%%%%%%%%%%%%%%%%%%%%%%%%%%%%%%%%%%%%%%%%%%%%%%%%%

%%%%%%%%%%%%%%%%% APPENDICES %%%%%%%%%%%%%%%%%%%%%
\appendix
\section{Time evolution of HOD -- more snapshots}
\label{appendix:hod_evol}

Here we include the ELG occupancies measured from the hydrodynamical simulations for all gravity models at 9 different redshifts. These are similar to the top row of Fig.~\ref{fig:HOD examples}, but the larger number of snapshots allows us to see the evolution sequence more clearly. 

\begin{figure*}
    \centering
    \includegraphics[width=15cm]{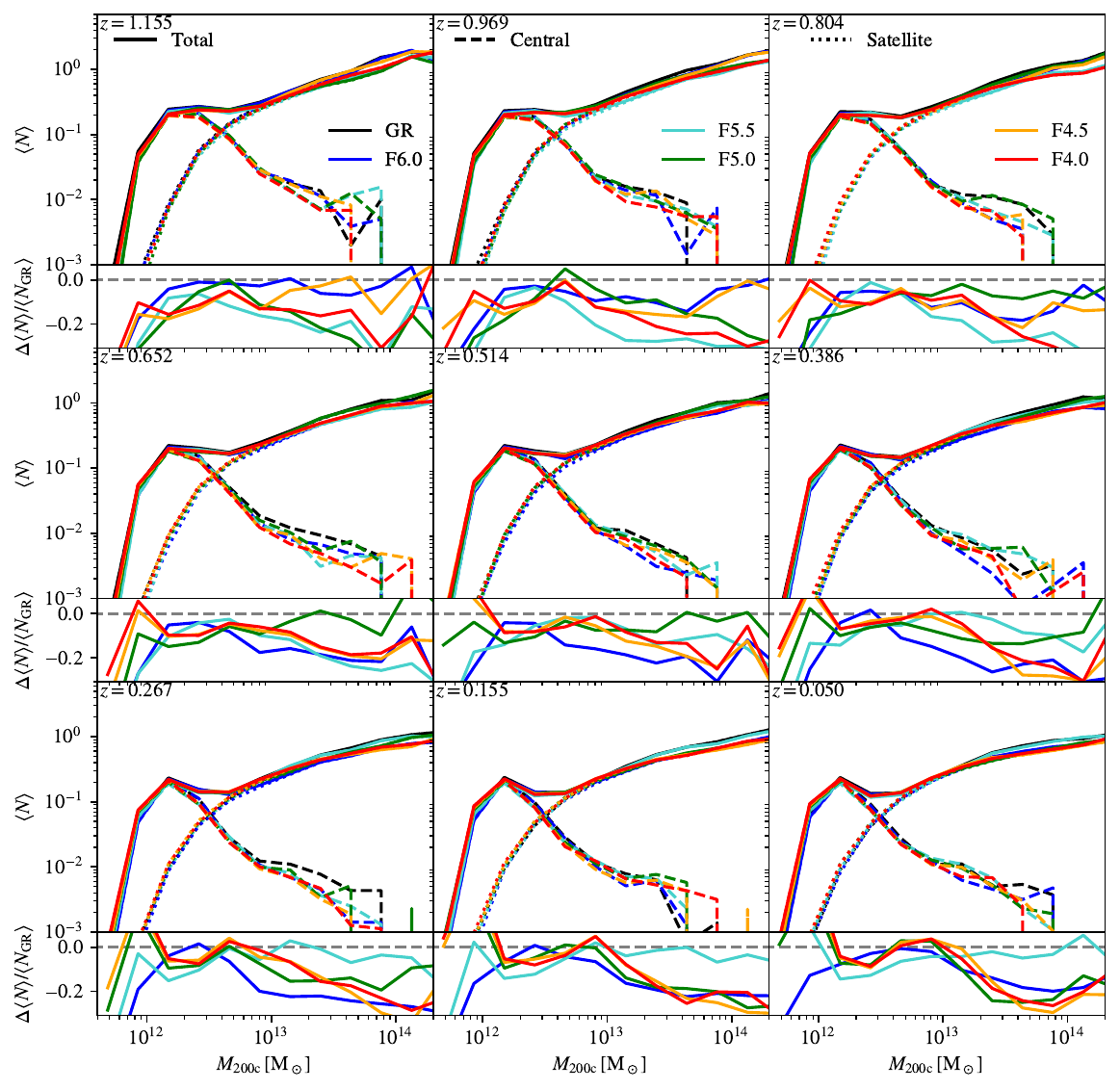}
    \caption{The same as the top row of Fig.~\ref{fig:HOD examples}, but with a larger number of redshift snapshots.}
    \label{fig:ELG_HOD_evol_more_snapshots}
\end{figure*}
%\input{appendix_environmental_dependence}

%%%%%%%%%%%%%%%%%%%%%%%%%%%%%%%%%%%%%%%%%%%%%%%%%%

% Don't change these lines
\bsp	% typesetting comment
\label{lastpage}
\end{document}